\documentclass[ aip, pof, amsmath,amssymb, reprint,onecolumn,superscriptaddress,showkeys]{revtex4-1}

\include{bookdefs}
\usepackage{graphicx}
\usepackage{amsmath,amsthm,amsfonts,amscd}

\renewcommand{\vec}[1]{\mbox{\boldmath$#1$}}

\usepackage{color}
\usepackage{rotating}
\definecolor{orange}{rgb}{.9,.3,0}

\usepackage{float}
\usepackage{calc}

\begin{document}

\title{Anomalous dispersion of Lagrangian particles in local regions of turbulent flows revealed by convex hull analysis}
                                            
\author{J. Pratt}
\email[]{jane.pratt@ipp.mpg.de}
\affiliation{Max-Planck-Institut f\"ur Plasmaphysik, 85748 Garching, Germany}
\affiliation{Max-Planck-Institut f\"ur Sonnensystemforschung, 37191 Katlenburg-Lindau, Germany}
\author{A. Busse}
\affiliation{School of Engineering, University of Glasgow, Glasgow G12 8QQ, United Kingdom}
\author{W.-C. M\"uller}
\affiliation{Center for Astronomy and Astrophysics, ER 3-2, TU Berlin, Hardenbergstr. 36, 10623 Berlin, Germany}

\author{S.C. Chapman}
\affiliation{Centre for Fusion, Space and Astrophysics, Physics Department, University of Warwick, Coventry, CV4 7AL, United Kingdom}
\affiliation{Max-Planck-Institut  f\"ur Physik komplexer Systeme, 01187 Dresden, Germany}
\affiliation{Department of Mathematics and Statistics, University of Tromso, Norway}

\author{N.W. Watkins}
\affiliation{Centre for Fusion, Space and Astrophysics, Physics Department, University of Warwick, Coventry, CV4 7AL, United Kingdom}
\affiliation{Max-Planck-Institut  f\"ur Physik komplexer Systeme, 01187 Dresden, Germany}
\affiliation{Centre for the Analysis of Time Series, London School of Economics, London, United Kingdom}
\affiliation{Department of Engineering and Innovation, Open University, Milton Keynes, United Kingdom}
\date{\today}

\begin{abstract}
 Local regions of anomalous particle dispersion, and intermittent events that occur in turbulent flows can  
 greatly influence the global statistical description of the flow. These local behaviors can be identified and analyzed by comparing the growth of 
neighboring convex hulls of Lagrangian tracer particles.   
Although in our simulations of homogeneous turbulence the convex hulls generally grow in size, after the Lagrangian particles that define the convex hulls begin to disperse, our analysis reveals short periods when the convex hulls of the Lagrangian particles shrink, evidence that particles are not dispersing simply. 
 Shrinkage 
 can be associated with anisotropic flows, since it occurs most frequently in the
  presence of a mean magnetic field or thermal convection.  
We compare dispersion between a wide range of statistically homogeneous and stationary turbulent flows
ranging from homogeneous isotropic Navier-Stokes turbulence
over different configurations of magnetohydrodynamic turbulence and Boussinesq convection.
\end{abstract}

\keywords{dispersion; intermittency; Lagrangian statistics; turbulence; MHD; convection}
\pacs{}

\maketitle

\vspace{5mm}

\section{Introduction}

Turbulent dispersion governs the spreading of contaminants in the environment, mixing of particles in combustion engines or in stellar interiors, accretion in proto-stellar molecular clouds, acceleration of cosmic rays,  and escape of hot particles from fusion machines.  Because of its general influence, characterization of particle dispersion in turbulent flows is of practical interest to physicists and engineers.     Here we address the broadly relevant cases of particle dispersion in statistically homogeneous magnetohydrodynamic (MHD) turbulence, Boussinesq \citep{glatzmaier2013introduction} convection, and Boussinesq MHD convection.  

The Lagrangian viewpoint is particularly suited to the investigation of the turbulent dispersion.  A Lagrangian treatment of turbulence provides a description of the paths of non-interacting fluid particles in a turbulent flow.  
    Conventionally, the relative dispersion of two, three, or four particles is used to characterize particle dispersion \citep{hackl2011multi,sawford2013gaussian,luthi2007lagrangian,toschibodenrev,xu2008evolution,lacasce2008statistics,yeungreview,pumir2000geometry,lin2013lagrangian}. 
When examining dispersion in anisotropic flows, calculation of the relative dispersion of particles requires time- and space-averaging  \citep{dubbeldam2009new} to produce statistically meaningful results.  These time- and space-averaged statistics can be difficult to interpret physically for an anisotropic flow or a flow containing evolving large-scale structures; simpler, more physically intuitive diagnostics are a helpful step.  In this work we develop an alternative to the standard Lagrangian multi-particle statistics that produces results close to a human perception of particle dispersion using a geometrical object called a convex hull \citep{efron65}. 

Convex hull analysis is in the same spirit as following a drop of dye as it spreads in a fluid, or following a puff of smoke as it spreads in the air, both classical fluid dynamics problems \citep{gifsmokepuffs,elder1959dispersion}.  As a diagnostic, the convex hull is distinguishable from two, three, or four particle Lagrangian statistics in that it uses many more particles.   Because of this, the convex hull is able to capture the extremes of the excursions of a group of particles, information relevant to the non-Gaussian aspects of the dynamics.   The results of convex hull analysis can be both more intuitive and more adaptable than two particle Lagrangian statistics.  By calculating convex hulls of groups of Lagrangian fluid particles, any geographic sub-region of a simulation volume can be analyzed and its diffusion properties can be
conveniently compared with other sub-regions.  This flexible local approach is particularly useful
when analyzing anisotropic flows 
that develop during vigorous convection \citep{shearbursts,mazzitelli2014pair,maeder2008convective,brun2011modeling,leprovost2006self}, when strong magnetic
fields are present \citep{beresnyak2006polarization,muller2003statistical,mason2006dynamic,esquivel2011velocity}, when magnetic reconnection is occurring \citep{eyink2011fast,yokoi2013transport}, or in turbulence simulations that have rare events or interfaces between turbulent and non-turbulent flows \citep{bifextreme,hunt2006mechanics}.  

Convex hulls of properly chosen particle groups deliver sufficiently well-converged statistics based on measurements at the positions of the particles that make up the hulls' surfaces. These statistics are localized in space, and at the same time associated with a spatial scale that is set by the characteristic extent of the hulls. In this light, the comparison between convex hull statistics
and two, three, or four particle Lagrangian dispersion statistics could be regarded as analogous to the comparison between wavelets
and Fourier transformations.

Recently, convex hull calculations have been used
to study diverse topics such as the size of spreading
GPS-enabled drifters moving on the surface of a lake
\citep{lakes}, star-forming clusters \citep{schmeja}, forest fires \citep{forestfire}, proteins \citep{li08,millett2013identifying}, or clusters of contaminant particles \citep{dietzel2013numerical}.
Studies of the relationships
between random walks, anomalous diffusion, extreme statistics and convex
hulls have been motivated by animal home ranges \citep{eco2006,ecoPRL,maj2010,vander2013trophic,dumonteil2013spatial}.  Convex hulls have also been used to study certain analytical statistics of Burgers turbulence by analogy with Brownian motion \citep{avellaneda1995statistical,bertoin2001some}.
In this work we consider dispersion, measured by the relative distance of particles, rather than diffusion, measured by the distance of a particle from its initial position. 
Our use of the convex hull to examine statistics of Lagrangian multi-particle dispersion is new. 
 Developments  to deploy fast algorithms for convex hulls on
hybrid architectures \citep{Stein2012265,Tang2012498}
may make convex hull calculations even more attractive as a diagnostic tool in the future.

\begin{figure}
\resizebox{6.5in}{!}{\includegraphics{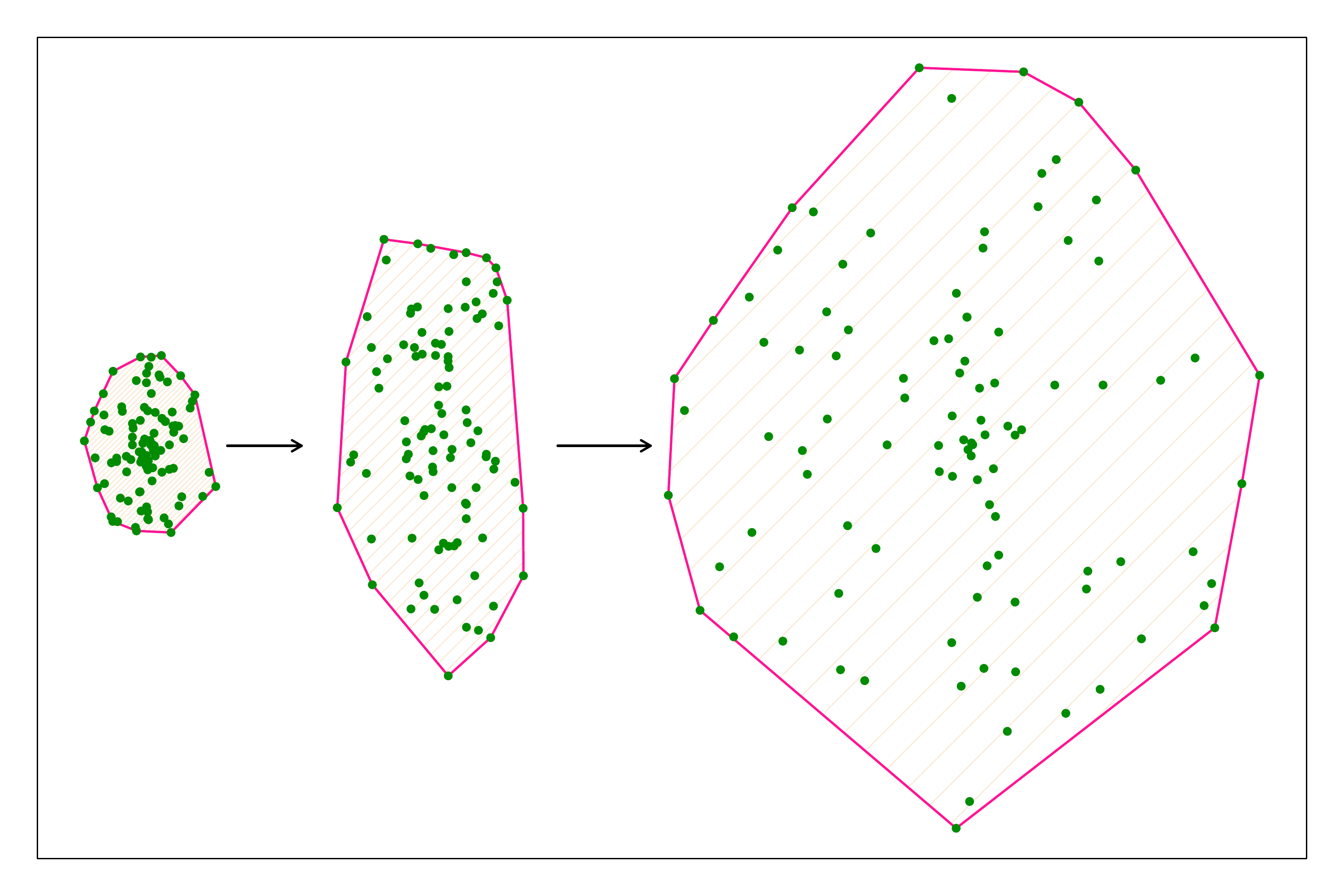}}
\caption{ An illustration of a two dimensional convex hull surrounding a group of particles (solid points) as they disperse in time.  The simulations we consider are incompressible, so the density of the fluid remains constant.  But the surface area and volume of the convex hull grow as the particles spread out from an initial volume.
\label{hull} }
\end{figure}

Conservation of volume is a primitive concept for mechanics of incompressible fluids.   A volume of fluid that is convex at an initial time will have the same volume after it evolves in a flow but the shape of the volume will generally not be a simple shape and it will no longer be convex.   The volume of the convex hull of the tracer particles that are enclosed in the initial volume will not be generally conserved at this later time.  This is illustrated in FIG.~\ref{hull}.  Surface area and volume growth are natural concepts for convex hulls.
We examine these concepts in a range of turbulent flows subject to different physics.
 
MHD turbulence \citep{busse_hoho,homann2013structures} and hydrodynamic convection \citep{schumacher09,schu2008,forster2007parameterization} are areas where statistical analysis of Lagrangian fluid particles has only recently begun to be applied.  This work presents new Lagrangian results from three-dimensional direct numerical  simulations of MHD convection, and compares them with results from MHD turbulence, hydrodynamic convection, and homogeneous isotropic turbulence.    This work is structured as follows.  In Section II we describe the fluid simulations and the method of calculating convex hulls of groups of Lagrangian fluid particles.  In Section III we present results showing that convex hulls may episodically shrink, and how different physical phenomena affect this.  In Section IV we discuss and summarize our results.

\section{Simulations}

We investigate the dispersion of Lagrangian particles in five different types of highly-turbulent systems: forced homogeneous isotropic Navier-Stokes turbulence
(simulation NST)\citep{sawfordreview,biferale05}, Boussinesq hydrodynamic convection (simulations HC1-5)\citep{chandrasekhar:book},
isotropic MHD turbulence with no mean magnetic field (simulation IMT)\citep{busse_hoho}, strongly anisotropic MHD turbulence with a 
high mean magnetic field $B_{z} \approx 10~ b_{\mathsf{rms}}$ relative to the root-mean-square
magnetic-field fluctuations (simulation AMT)\citep{angeladiss09}, and Boussinesq MHD convection simulations (simulations MC1-10)\citep{shearbursts, moll2011}. 
The range of convective flow behaviors is wide and includes intermittent events like shear bursts \citep{shearbursts}, and the formation of large-scale magnetic structures on long intervals.  For this reason we must examine multiple simulations of hydrodynamic convection and MHD convection with different initial conditions and different fundamental parameters in order to draw broadly relevant conclusions. 

In each direct numerical simulation, nondimensional equations are solved using a pseudospectral
method.  In most simulations the volume is a cube with a side of length $2\pi$.  In some of the MHD convection simulations (MC1, MC5, MC9, MC10) the volume is a slab that has a slightly larger extent of $2^{3/2} \pi$ in the $x$- and $y$-directions.  The non-dimensional Boussinesq equations for MHD convection are
\begin{eqnarray} \label{realbmhdc}
\frac{\partial \vec{\omega} }{\partial t} &-& \nabla \times (\vec{v} \times \vec{\omega} +  \vec{j} \times \vec{B}) 
=  \hat{\nu} \nabla^2 \vec{\omega} - \nabla \theta \times \vec{g}_0
\\
\frac{\partial \vec{B} }{\partial t} &-& \nabla \times (\vec{v} \times \vec{B}) =  \hat{\eta} \nabla^2 \vec{B}
\\ \label{thermeq}
\frac{\partial \theta }{\partial t} &+& (\vec{v} \cdot \nabla) \theta = \hat{\kappa} \nabla^2 \theta -  (\vec{v} \cdot \nabla) T_0\\
&& \nabla \cdot \vec{v}=  \nabla \cdot \vec{B}=0 ~~.
\end{eqnarray}
These equations include the solenoidal velocity field $\vec{v}$,
vorticity $\vec{\omega}=\nabla\times\vec{v}$, magnetic field $\vec{B}$, and 
current $\vec{j}=\nabla\times\vec{B}$.   The
magnetic field is given in Alfv\'enic units, with an Alfv\'en Mach number
$v_0/v_{\mathrm A} = 1$, where $v_0$ is defined by the characteristic length and time scales of the convective motions.  
  The quantity $\theta$ denotes the temperature fluctuation about
a linear mean temperature profile $T_0(z)$ where $z$ is the direction of
gravity.  In eq. \ref{thermeq} this mean temperature provides the convective drive of the system.  In eq. \ref{realbmhdc}, the term including the temperature fluctuation $\theta$ is the buoyancy force.  The vector $\vec{g}_0$ is a unit vector in the direction of gravity.  Three dimensionless parameters
appear in the equations: $\hat{\nu}$, $\hat{\eta}$, and $\hat{\kappa}$.
They derive from the kinematic viscosity $\nu$, the magnetic diffusivity
$\eta$, and thermal diffusivity $\kappa$.   For simulations of hydrodynamic convection, the magnetic field $\vec{B}$ is set to zero.  For simulations of MHD turbulence,  the mean temperature gradient $T_0(z)$ and thermal fluctuations $\theta$  are zero.  For simulations of hydrodynamic Navier-Stokes turbulence, both magnetic field terms and temperature terms are zero.  A fixed time step and a trapezoidal leapfrog method \citep{kurihara1965use} are used to integrate the equations for simulations NST, IMT, AMT, and HC1.  The remaining Boussinesq convection simulations are integrated in time with a low-storage 3rd-order Runge Kutta time scheme \citep{will80} and an adaptive time step, which allows for better time resolution of instabilities that occur during MHD convection.

 A summary of each simulation is given in Table~\ref{simsuma}. 
In this table, we define the Reynolds number to be $\mathsf{Re= \langle E_v^{1/2} L\rangle /\hat{\nu}} $, where $\mathsf{E_v}= \vec{v}^2/ 2$ is the kinetic energy, and the brackets indicate a time-average. 
We define the characteristic length scale $\mathsf{L}$ based on the largest-scale motions of the system in question.
 For statistically homogeneous turbulent convection the characteristic length scale is the instantaneous temperature gradient length scale $\mathsf{L}=T_*/\nabla T_0$ where $T_{*}$ is the root-mean-square of
temperature fluctuations and $\nabla T_0$ is the constant vertical mean temperature gradient.  For non-convective statistically homogeneous turbulent flows, the characteristic length scale is a dimensional estimate of the size of the largest eddies, $\mathsf{L}=\mathsf{E_v}^{3/2}/\mathsf{\epsilon_{\mathrm{v}}}$, where $\mathsf{\epsilon_{\mathrm{v}}}=\hat{\nu} \langle \sum_k k^2 \vec{v}^2 \rangle$ is the time-averaged rate
of kinetic energy dissipation.
 The magnetic Reynolds number is defined from the Reynolds number and the magnetic Prandtl number, i.e. $\mathsf{Re_m}=\mathsf{Pr_m Re}$. 
We measure length in units of the
Kolmogorov microscale $\mathsf{\eta_{kol}=(\hat{\nu}^3/\mathsf{\epsilon_{\mathrm{v}}})^{1/4}}$ and time 
in units of the Kolmogorov time-scale
$\mathsf{\tau_{\eta}=(\hat{\nu}/\mathsf{\epsilon_{\mathrm{v}}})^{1/2}}$; these are the smallest length and time scales that characterize turbulent flows.
The Kolmogorov microscale multiplied by $\mathsf{k_{\mathrm{max}}}$, the highest
wavenumber in the simulation, is often used to test whether a simulation
is adequately resolved on small spatial scales.  In this work simulations of different
resolution are investigated; all of the simulations fulfill the standard
criterion ($\mathsf{k_{\mathrm{max}} \eta_{kol} >1.5 }$) for adequate spatial resolution\citep{pope2000turbulent}.

\begin{table*}
\caption{Simulation parameters\label{simsuma}
 }
\begin{tabular}{lccccccccccccccccccccccccccc}
\hline\hline
                            & NST     & IMT     &  AMT               & HC1  & HC2        &  HC3     & HC4     &  HC5   &  MC1      &  MC2    &  MC3         & MC4    & MC5    & MC6  & MC7    & MC8  & MC9 & MC10
\\ \hline
 grid               & $1024$  & $512$& $512$ &$2048$ & $512$  & $512$  & $512$ & $512$  & $512$   & $512$ & $512$  & $512$ & $512$ & $512$ & $512$ & $512$ & $512$ & $512$
 \\ \hline
 $\mathsf{n_p}$ ($10^6$)   & 3.2  & 0.5 &  0.5 &12.5 & 1.0  & 1.0  & 1.0 &1.0  & 1.0   & 1.0 & 1.0  & 1.0 & 1.0 & 1.0 & 1.0 & 1.0 & 1.0 & 1.0
\\ \hline
$\mathsf{Re}$ ($10^3$) & 2.9   & 1.4    & 3.5                 & 6.4 &  1.7       &  3.6    & 2.0    & 1.5   &   3.1   & 5.2     & 5.4      & 5.6     &   3.5   &  5.1  & 4.0 & 2.4   & 2.4 & 1.3
\\ \hline 
$\mathsf{Re_m}$ ($10^3$)                     & -  & 1.4 & 3.5 & - & - & - & - & -  & 6.2  & 5.2 & 5.4 & 2.8 & 7.0 & 7.65 & 8.0 & 4.2 & 4.8 & 3.9
\\ \hline
$\mathsf{Pr}$               & -       & -      &  -                  &  0.6 &  1.0       &  1.25    & 2.0     &  2.0 &   1       & 1       & 0.5     &   1     &  1      &  2    & 1.5  & 1.76  & 1.3 & 1
\\ \hline
$\mathsf{Pr_m}$             & -       &  1      &   1                & -    &  -         &  -       &  -      &  -   &   2       & 1       & 1        &    0.5  &  2      & 1.5   & 2    & 1.76  & 2.0 & 3
\\ \hline 
 $\eta_{\mathsf{kol}}$ ($10^{-3}$)  & 4.58    & 9.72    & 9.94               & 1.63  & 7.26      & 7.18     & 6.97    & 12.6 & 7.88     &  7.39    & 6.29     &  6.7    &  7.8    & 8.9   & 9.2  & 12.5  & 7.8 & 16.1
\\ \hline
$\tau_{\eta}$($10^{-2}$) & 5.25 & 9.44    & 14.1              & 1.49  & 2.64      & 2.58     & 2.43    & 3.97  &  3.11 & 2.73    & 2.64     & 2.98   &  3.02   &  2.6  & 2.8  & 3.6   & 3.04 & 4.23
\\ \hline
LCT ($\tau_{\eta})$ & 276.1 & 275.3  & 378.3                 & 538.1 & 648.8    & 727.1   & 624.8  & 339.4         & 1219.2 & 529.9 & 741.7 & 685.3  &  431.8 & 807.7 & 691.4 & 581.0 &  825.4 & 494.0
\\  \hline
$r_{\mathrm{A}}$                         & -    & 0.56    & 0.78        & -  & -      & -     & -    & -                             & 1.28 & 2.20 & 1.43 & 2.79  &  1.12 & 1.78 & 1.72 & 2.00 &  1.12 & 1.30
 \\  \hline
  $\bar{\ell}_{\mathrm{bo}}$     & -    & -    & -        & 0.21  & 0.27    & 0.27     & 0.28  & 0.25          & 0.08 & 0.09 & 0.08 & 0.06  &  0.09 & 0.12 & 0.12 & 0.12 &  0.08 & 0.13
 \\ \hline
 $\mathsf{n_{pch}}$      & 24       & 51     &  60     & 88         & 90        & 61      & 59       & 147       & 60       & 61         & 61           & 61        & 61       & 52       & 380,  61     & 55        & 60     & 54
  \\ \hline
 $\mathsf{N_{hulls}}$   & 5000  & 2500 &  2500 &14946  & 1200   & 2640  & 720    &  4420   & 720      & 720      & 720         & 720     & 2160   & 3700 & 2500, 720   & 720      & 2400  & 3940
  \\  \hline
 $\ell_{\mathrm{hull}}$ ($\eta_{\mathsf{kol}}$)  & 27   & 25   & 25  & 38  & 34     & 34   & 35   & 19   & 31        & 33        & 39            & 37        &  32      & 39      & 40, 27       & 20       &  31      & 15
\\ \hline\hline
\end{tabular}
\end{table*}

Formulation of optimal boundary conditions for simulations of turbulent flows is 
delicate because boundaries strongly influence
the structure and dynamics of the flow.  For homogeneous isotropic turbulence, it is standard to employ boundary conditions that are periodic in $x$, $y$, and $z$.  These fully-periodic boundary conditions are used for simulations NST, IMT, AMT, and HC1.  In simulation HC1, the temperature gradient is perpendicular to the direction of gravity. 
 For the remaining convection simulations the choice of fully-periodic boundary conditions (sometimes called
  homogeneous Rayleigh-B\'enard boundary conditions when applied to convection) allows macroscopic elevator instabilities to form \citep{calzavarini_etal:elevator}.  These instabilities destroy the natural pattern of the original turbulent flow field.
For Boussinesq convection, the simulation volume considered in this work is confined by quasi-periodic boundary conditions.  
   The only additional constraint in the quasi-periodic boundary conditions is the explicit suppression of mean flows
parallel to gravity, which are removed at each time step.
 Our simulations are pseudospectral, and therefore the mean flow is straightforwardly isolated as the $k=0$ mode in Fourier space, which corresponds to the volume-averaged velocity.
  Quasi-periodic boundary conditions combine the conceptual simplicity of statistical homogeneity with a
 physically natural convective driving of the turbulent flow.  These boundary conditions do not enforce a
large-scale structuring of the turbulent flow, such as the convection-cell
pattern caused by Rayleigh-B\'enard boundary conditions.  In the quasi-periodic simulations presented in this work, we find no evidence of the macroscopic elevator instability
 although we follow the evolution of the flow for long times. 
   Quasi-periodic boundary conditions allow for close comparison with simulations that use fully-periodic boundary conditions.

Simulations NST, IMT, and AMT are forced using Ornstein-Uhlenbeck processes with a finite time-correlation on the order of the autocorrelation time of the velocity field (for details of this forcing method, see \citep{angeladiss09}).
 The convection simulations are Boussinesq systems driven solely by a constant temperature gradient in the vertical direction, opposing the direction of gravity.  The magnetic fields present in our MHD convection simulations are generated self-consistently by the flow from a small seed field through small-scale dynamo action.   The magnetic fields are allowed to evolve through the kinematic phase of the dynamo until a steady-state is reached.
 For Boussinesq convection, a length-scale that characterizes the scale-dependent importance of convective driving is the Bolgiano-Obukhov length, {$\ell_{\mathrm{bo}}=\mathsf{\epsilon_{\mathrm{v}}^{5/4}/\epsilon_{\mathrm{T}}^{3/4}}$},  where  $\mathsf{\epsilon_{\mathrm{ T}}}$ is the average rate of thermal energy dissipation. 
  This length scale separates convectively-driven scales of the flow {$\ell > \ell_{\mathrm{bo}}$} from the range of scales where the temperature fluctuations behave as a passive scalar $\ell < \ell_{\mathrm{bo}}$.    In Table~\ref{simsuma} this length scale is averaged over the simulation time, normalized to the height of the simulation volume, and recorded as $\bar{\ell}_{\mathrm{bo}}$.  The table also includes the mean Alfv\'en ratio, $r_{\mathrm{A}} = \langle \mathsf{E_v} /\mathsf{E_b} \rangle$, the time-average of the kinetic energy divided by the magnetic energy $\mathsf{E_b}= \vec{B}^2 / 2$. 

The positions of Lagrangian fluid particles (i.e. passive tracers) are initialized in a homogeneous random distribution at a time when the turbulent flow is in a steady state. The total number of particles in the simulation,  $\mathsf{n_p}$, is listed in Table~\ref{simsuma}.  At each time step the particle velocities are interpolated from the instantaneous Eulerian velocity field using either a trilinear (for simulations HC1-5, and MC1-10) or tricubic (for simulations NST, IMT, and AMT) polynomial interpolation scheme.  Particle positions are calculated by numerical integration of the equation of motion using a predictor-corrector method.  
Each simulation is run for a sufficient time that Lagrangian particle pairs have separated, on average,
 by the length of the simulation volume.   This time is the Lagrangian crossing time (LCT) listed in the table in units of the Kolmogorov time scale.   
  Lagrangian statistics exhibit a diffusive trend near this time.

To calculate convex hulls, we select groups of Lagrangian particles initially contained in small sub-volumes of each simulation that have the same shape and proportions as the total simulation volume.  We follow the three-dimensional convex hull, the smallest convex three-dimensional polygon that encloses each group of particles, for the span of the simulation.  The Lagrangian particle data is sampled at a rate of 10 time steps for simulations NST, IMT, AMT, and HC1,  and at a rate of 40 time steps for the remaining convection simulations.  Selection of particle groups based on initial position yields groups of particles that contain
nearly the same number of particles, but some variation occurs based on the grid-size of the simulation, the total number of particles, and the initial random distribution of the particles. 
The average number of particles that define a convex hull,  $\mathsf{n_{pch}}$, is listed for each simulation in Table~\ref{simsuma}.
  To calculate the convex hull of groups of particles, we use the standard QuickHull algorithm\citep{Barber96thequickhull,2013barber}, implemented in the function \emph{convhulln} in the package \emph{geometry} publicly available for R \citep{ihaka1996r,cranr}.
The surface area and volume of the convex hulls are calculated as part of this package at every time using a Delaunay triangulation.

For non-convective systems, the ensemble of convex hulls is initially selected to fill completely a horizontal slab.  For convection simulations, multiple horizontal slabs, placed at heights spread systematically throughout the simulation volume, are treated.  The total number of convex hulls that we track is listed as $\mathsf{N_{hulls}}$ in Table~\ref{simsuma}, and is substantially more than required for statistical convergence of the average quantities that we calculate in this work.  This large number of convex hulls is examined in order to accurately include statistically less common flow features.

\subsection{Verification of convex hull calculations}

As any pair of particles separates in a turbulent flow, the particles move with the small-scale fluctuations of the velocity field. 
Sometimes the particles separate, and sometimes they move closer together. The distance between the two particles generally
increases in time, but with an erratic, noisy signature.  If a convex hull is defined by a very small group of particles, then most of the particles
are used to define the surface of the convex hull, and are called vertices of the convex hull.  In this situation, the convex hull, like the particle-pair distance, shrinks or grows erratically as its component particles move in the turbulent flow.  However for a convex hull that is defined by a large group of particles, vertices of the convex hull that move toward the center of the group can easily become interior particles of the convex hull rather than vertices.   Typically the particles
that are vertices of the convex hull are exchanged frequently.  If any particular
vertex moves inward toward the center of the group of particles, it is
unlikely that it will remain a vertex of the convex hull because of the required convexity of the surface.  Other nearby particles will continue to move away from the group,
and the convex hull will continue to expand smoothly.   Thus having a sufficient number of particles inside each convex hull is important to eliminate physically uninteresting noise from the convex hull evolution.

To test optimum particle-group sizes, preliminary to convex hull calculations we performed a study of the effect of the size of the initial convex hulls.  For initial convex hulls where the length of the side spans 5\% of the simulation box-length, a ballistic regime is no longer easily identifiable in the convex hull growth, and small-scale behaviors are obfuscated in the convex hull results. When initial convex hulls with a side less of 1\% or less of the simulation box-length are used, the convex hulls grow erratically because the number of particles that comprise the hull is too small. 
 We determine that an optimal length for the side of the initial convex hulls in our simulations is between 2\% and 4\% of the simulation box-length.  The side of the initial convex hulls,  $\ell_{\mathrm{hull}}$, is listed in Table~\ref{simsuma} for each simulation in units of the Kolmogorov microscale.  The size of the initial hulls that we are able to track is highly dependent on the density of particles tracked in our simulations; tracking significantly larger numbers of particles would allow us to examine smaller initial convex hulls.
In addition, noise from fluctuations that might appear on the time-scale of a single time-step of the simulation does not affect the convex hull calculations because these calculates sample the Lagrangian data at a rate of 10-40 time steps.

\begin{figure}
\resizebox{3.25in}{!}{\includegraphics{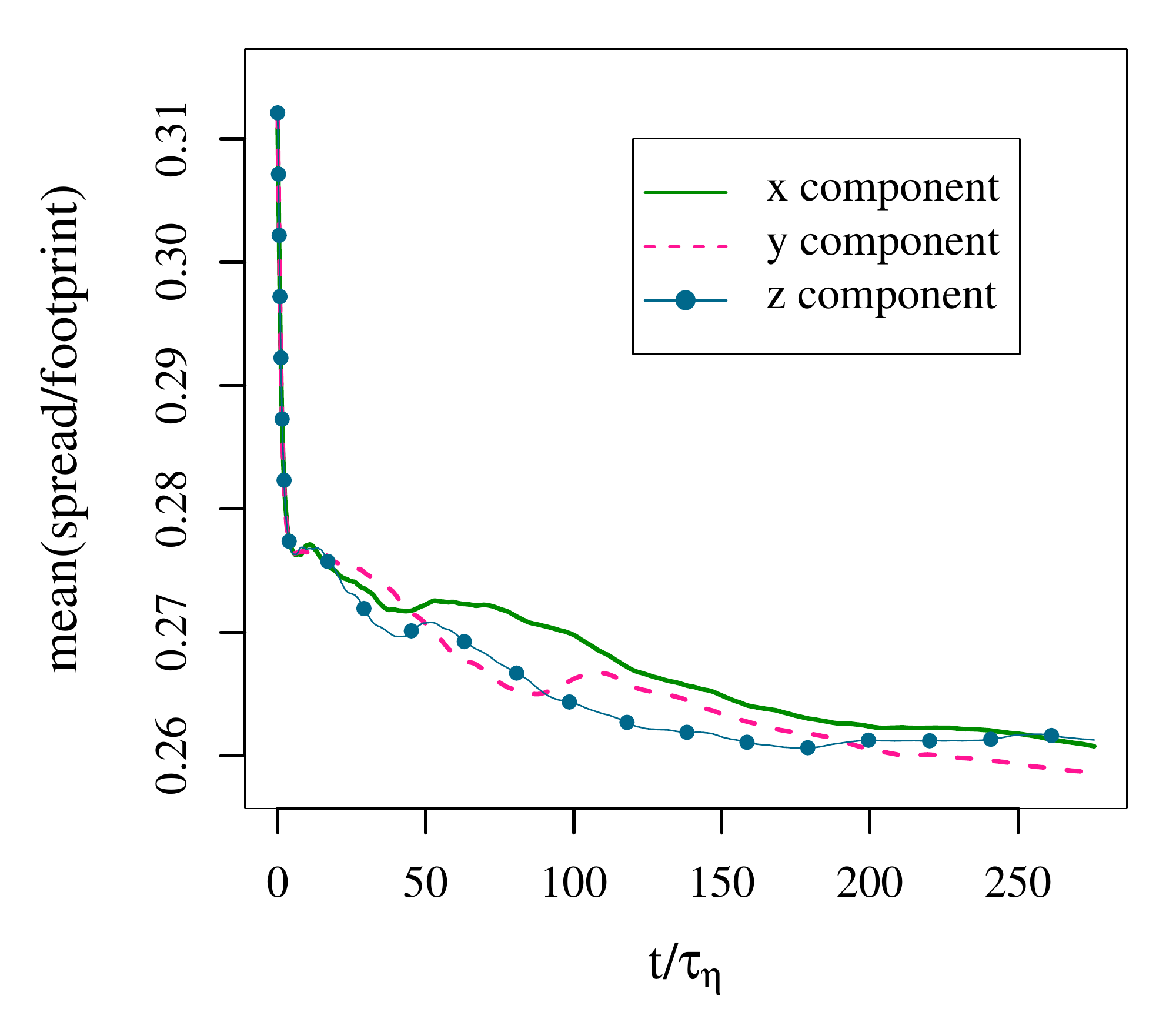}}\resizebox{3.25in}{!}{\includegraphics{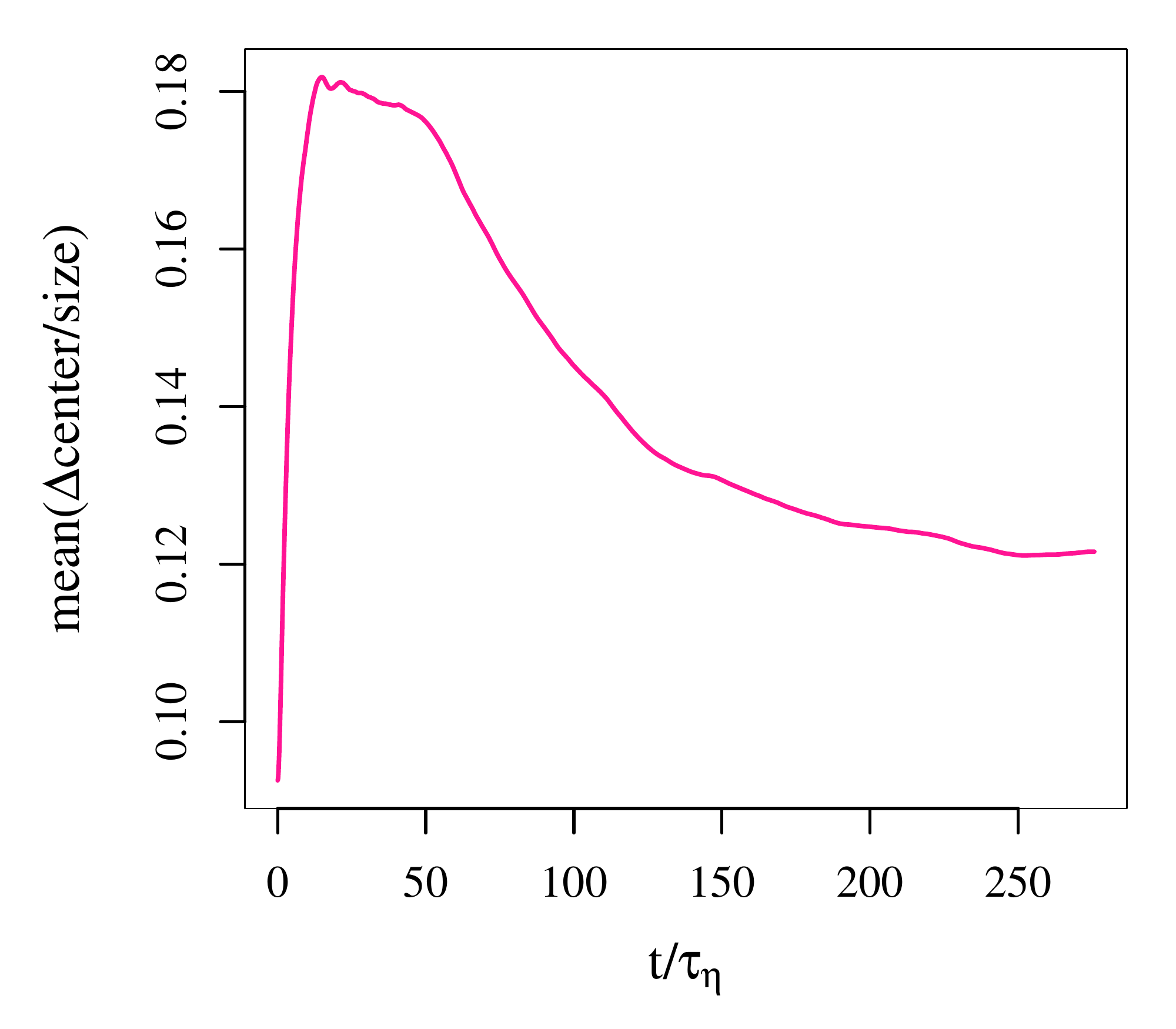}}
\caption{ 
(Left) standard deviation of the distribution of particle-positions for the particles contained in a convex hull, divided by the footprint of the hull in each direction, averaged over  all convex hulls in simulation NST. The footprint is the projection of the hull onto the $x$, $y$, or $z$ axis. These curves should become small if the particles do not spread through the interior of the convex hull as the convex hull expands.
(Right) the difference between the geometric center of the convex hull and the center of mass of the group of particles contained in the convex hull, divided by a measure of convex hull size, i.e. the square-root of the sum of the squares of the footprints in $x$, $y$, and $z$, averaged over all convex hulls in simulation NST.  This curve should become large if the particles contained in the convex hull do not move into
the space inside of the convex hull evenly as it grows. 
\label{newclumpgraph} }
\end{figure}

We stop tracking the convex hulls when the Lagrangian crossing time is reached.   By the Lagrangian crossing time, the convex hulls in the system have grown, on average, to fill a volume on the order of the simulation volume.
 A convex hull is defined by the particles within it that disperse the fastest.  
  As a convex hull evolves, particles inside the hull can move to the outer surface of the hull and become convex hull vertices.  Likewise particles that are convex hull vertices can move toward the inner domain of the convex hull and become interior particles rather than vertices.  The number of particles that are vertices of the convex hull generally decreases mildly with time;  this decrease is typically on the order of 10\% before the Lagrangian crossing time is reached, and happens gradually after the initial ballistic phase.   On average, the groups of particles remain 
  well-distributed and well-centered inside of the convex hulls throughout our simulations, even for highly-anisotropic systems.  This is illustrated for the Navier-Stokes turbulence simulation NST in FIG.~\ref{newclumpgraph}.  
After an initial ballistic phase of growth, the difference between the geometric center of the convex hull and the center of mass of the group of particles contained in the convex hull remains less than the standard deviation of particle positions.
  We conclude that the convex hull provides a reasonable description of the dispersion of the entire group of particles throughout the simulations that we examine in this work. 

\begin{figure}
\resizebox{6.5in}{!}{\includegraphics{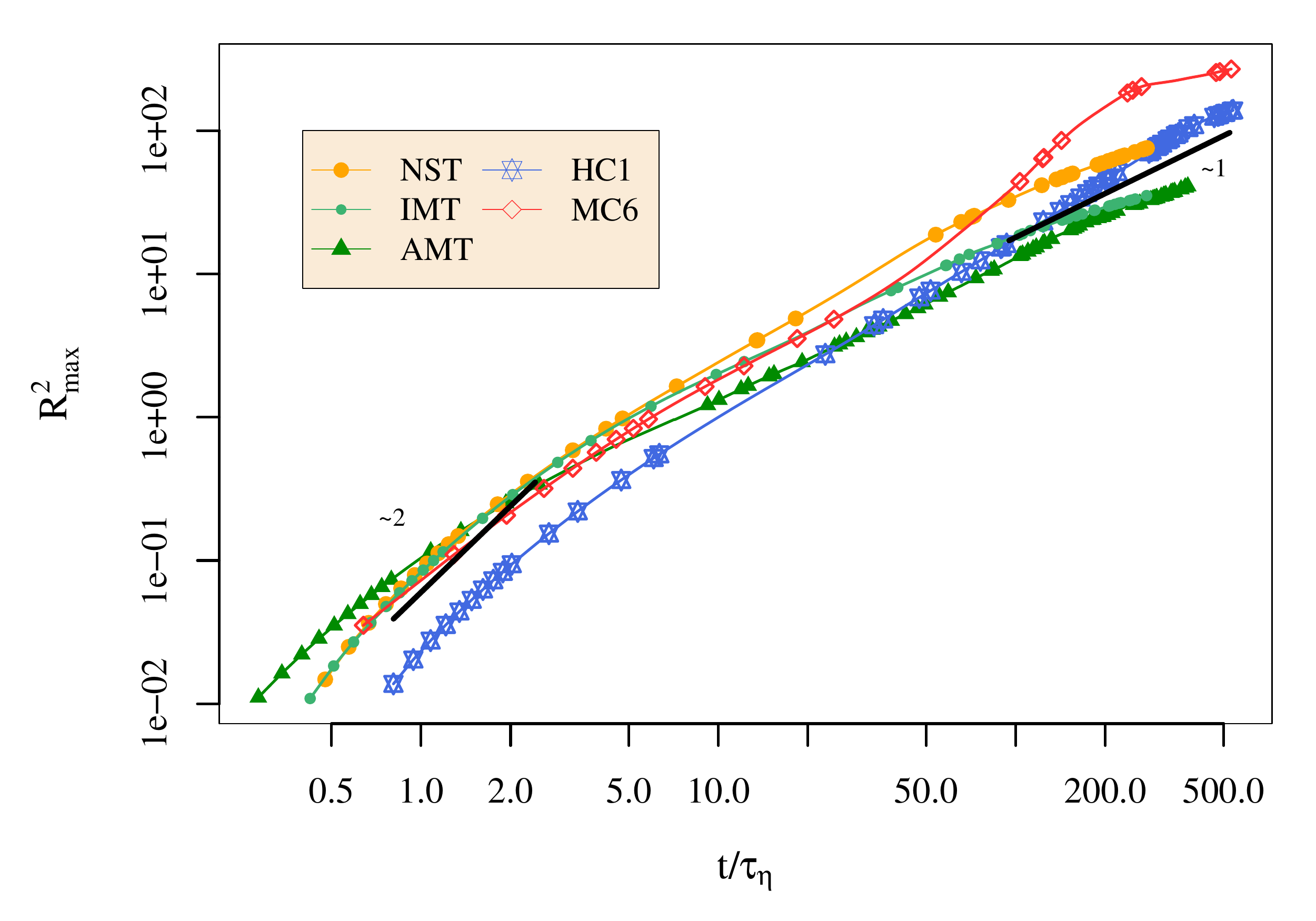}}
\caption{The average length of a maximal internal ray squared for an ensemble of evenly-spaced, local convex hulls examined in each system.
The length of the ray is normalized by its initial value and then shifted so that it is zero at the initial time, i.e. $R^2_{\mathsf{max}}=\langle r^2 \rangle / \langle r(t=0)^2 \rangle -1$ where $r$ is the maximum internal ray of a single convex hull, and the brackets represent an average over the ensemble of convex hulls.  Results from representatives of each of the 5 types of simulations described in Table~\ref{simsuma} are shown.
\label{eaxis} }
\end{figure}

In order to verify the results of our convex hull calculations we calculate a maximal internal ray of the convex hull.  If a convex hull were a perfect sphere, the maximal internal ray would be identical to the diameter of the sphere.   The results of averaging the square of the maximal internal ray over an ensemble of convex hulls in each system are shown in FIG.~\ref{eaxis}.   After evolving for some time, a convex hull grows significantly larger than its initial size, so the particles can be considered to have approximately the same origin point.   Therefore the maximal internal ray of the convex hull exhibits ballistic scaling at short times and diffusive scaling at long times comparable to Lagrangian single-particle diffusion.   FIG.~\ref{eaxis}  demonstrates that the maximal internal ray squared reproduces clear ballistic regimes with a slope of 2 at early times for all of the types of systems considered.  
Simulations where the turbulence is randomly forced (NST, IMT, and AMT) exhibit the expected diffusive regime with slope 1 at long times.  Large upwardly-moving or downwardly-moving flows force the convection simulations into a mildly super-diffusive or sub-diffusive regime with a slope that deviates slightly from 1.  Simulations HC1 and MC6 are good examples of this typical behavior associated with convection; both of these systems oscillate between super-diffusive and sub-diffusive slopes during the diffusive regime.  System-wide averages of convex hull surface area and volume show similar diffusive behaviors.

\section{Results}
\subsection{Local dispersive behaviors revealed by convex hull analysis}

\begin{figure}
\resizebox{3.25in}{!}{\includegraphics{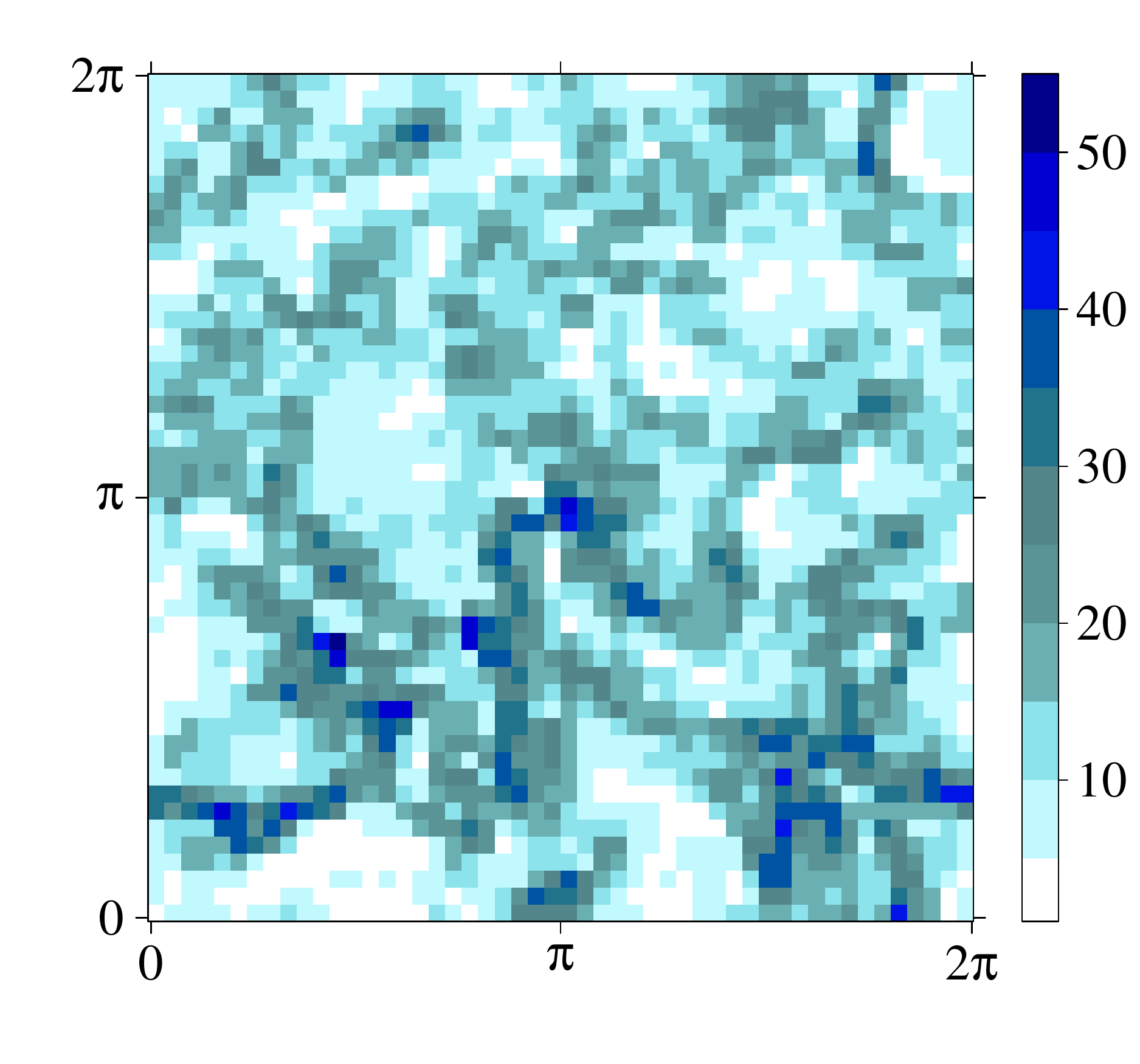}}\resizebox{3.25in}{!}{\includegraphics{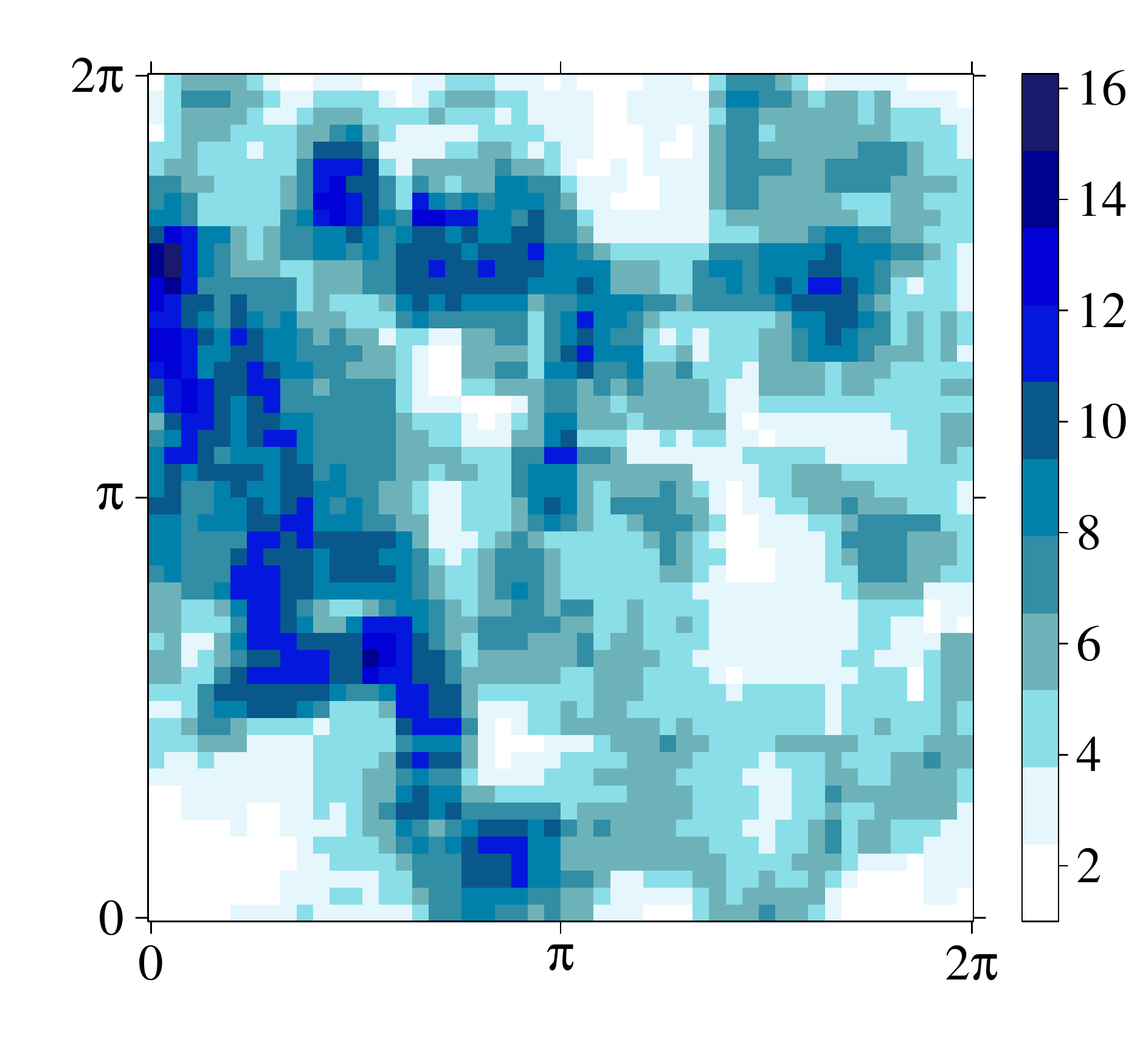}}
\caption{A contour plot showing the volumes of local convex hulls across an $x$-$y$ planar cut of the simulation volume from (left) the MHD simulation with strong mean magnetic-field AMT and (right) the MHD convection simulation MC7.  Volume is normalized to the minimum volume over convex hulls in the plane at this instant in time.  Color is indicated at the initial position of the convex hull.   The planar cut is in the direction perpendicular to the mean magnetic field for AMT, and perpendicular to gravity for MC7.  2500 local convex hulls fill the slab centered on $z=0.27$ (AMT), and $z=0.13$ (MC7) where the total height of the simulation box is $2 \pi$.  These snapshots were selected mid-way through each simulation, and are representative of long-time trends typical for these systems.
\label{volcontourplot} }
\end{figure}

Dispersion of particles in different sub-volumes of a flow can be faster or slower than the average.  In homogeneous isotropic flows the variation tends to even-out over long times, and averaging over initial conditions produces meaningful statistics.  \citet{hackl2011multi} note that in homogeneous isotropic turbulence, highly anisotropic dispersion of particles occurs but is generally short-lived.  We observe that when the flow is structured by a strong magnetic field or by convection, longer-lived regions of highly anisotropic particle dispersion arise, leading to super-dispersion in a particular direction.  These  regions of super-dispersion are dependent on the initial flow condition.  The sizes of convex hulls in these regions tend to greatly exceed the size of neighboring hulls, and continue to grow faster for the duration of the simulation.  The left panel of FIG.~\ref{volcontourplot} shows a characteristic time-snapshot of the volume of convex hulls in a planar cut of the MHD simulation AMT, which is anisotropic because of the presence of a strong mean magnetic field pointing into the plane.  At the instant in time pictured, roughly one third of the way through the simulation, some hulls in the ensemble have grown to be an order of magnitude larger than their neighbors.  
A similarly strong inhomogeneity in particle dispersion can result during vigorous convection, as shown in the right panel of FIG.~\ref{volcontourplot}.  This snapshot from the  MHD convection simulation MC7 was taken about half-way through the simulation time.  The size of super-diffusive structures that arise during convection is generally larger than in the stochastically forced turbulence simulations.

\begin{figure}
\resizebox{3.25in}{!}{\includegraphics{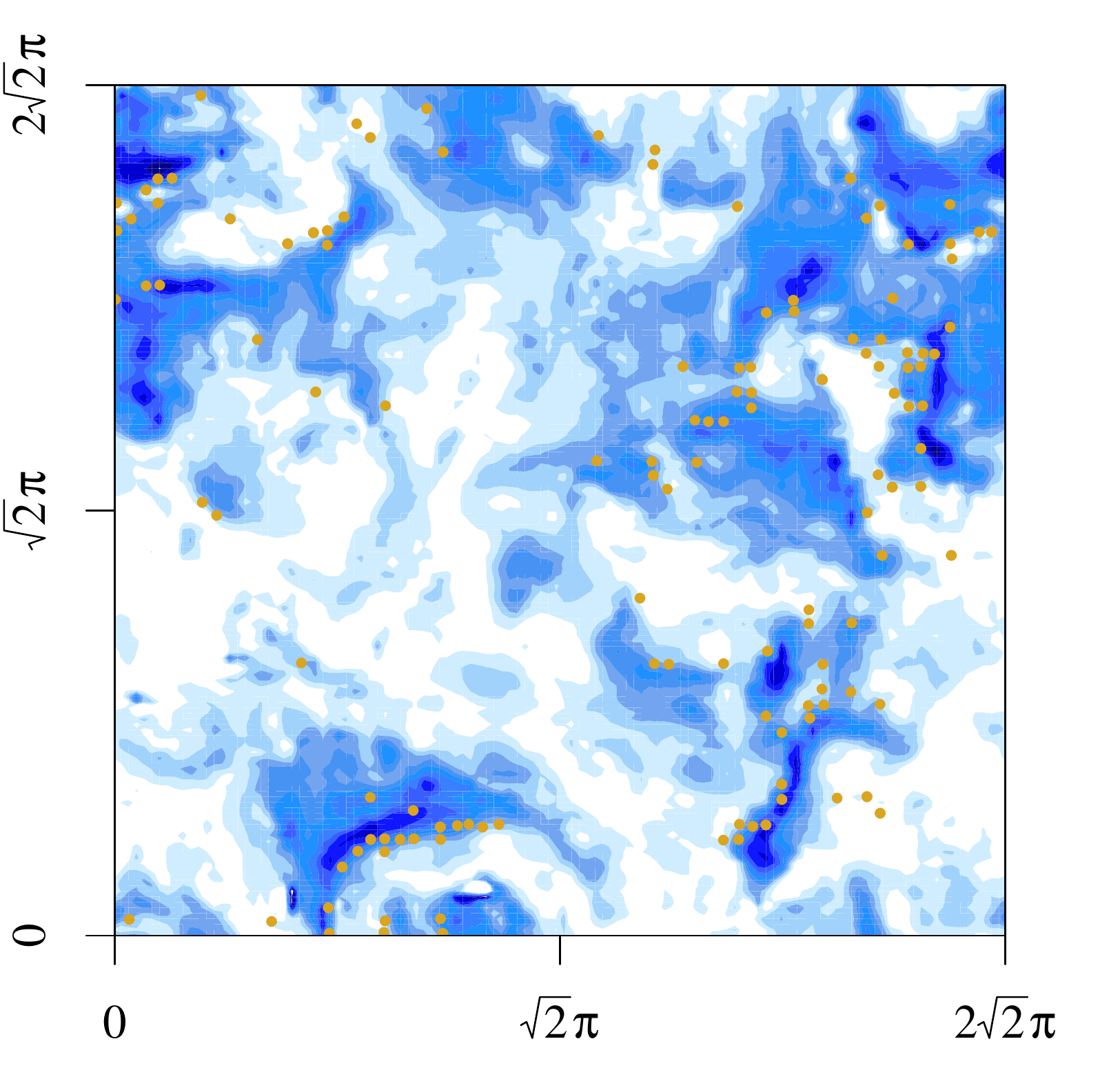}}\resizebox{3.25in}{!}{\includegraphics{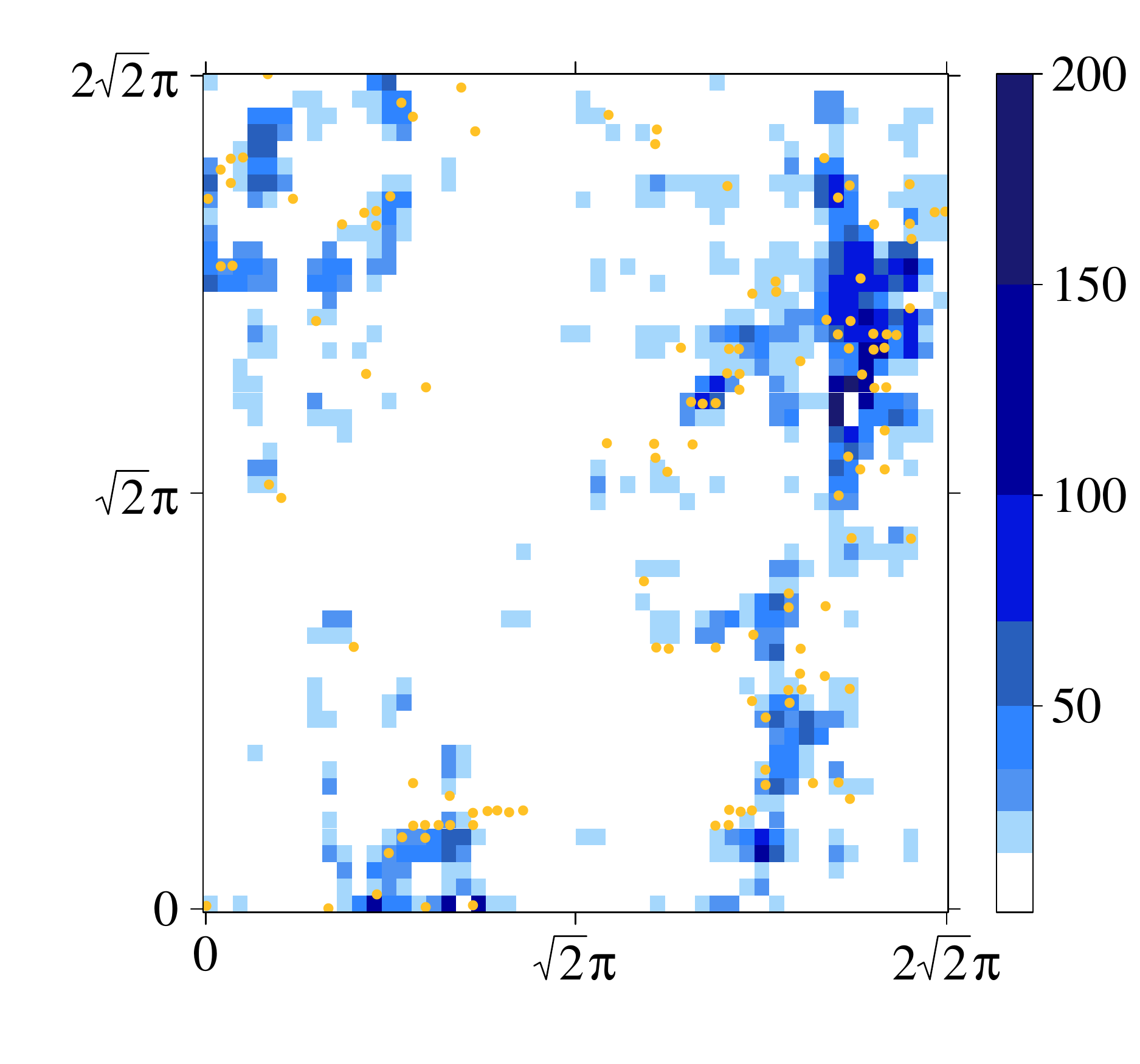}}
\caption{(Left) values of the finite-time Lyapunov exponent field (yellow dots) above an arbitrarily-selected high threshold, plotted over a contour plot of the initial velocity magnitude in simulation MC10. Darker colors indicate higher velocity magnitude.  (Right) values of the finite-time Lyapunov exponent field above an arbitrarily-selected high threshold, plotted over a contour plot of convex hull volume, normalized to the minimum volume over convex hulls in the plane at this instant in time.  This contour plot is created at an early point in time in the simulation, so that the LCS calculation is relevant to the dispersion behavior revealed by the convex hull volumes.
\label{lcscontourplot}
\label{lcsplot} }
\end{figure}

As  FIG.~\ref{volcontourplot} demonstrates, convex hull volumes can reveal neighboring regions that display different dispersive behaviors.
A mathematically rigorous tool to determine the boundaries of such regions has been developed over the last decade: Lagrangian coherent structures (LCS) \citep{haller2000lagrangian,shadden2005definition,karrasch2013finite}.   
 LCS are calculated from ridges in the finite-time Lyapunov exponent (FTLE) field, a measure of how much each pair of particles in
 the system separates after a fixed interval of time.  LCS calculations have been applied to laminar flows and flows with stationary structures.  In these studies, the LCS clearly outline areas with different diffusive behaviors, for example the boundaries between a vortex and the remaining flow field \citep{joseph2002relation, green2007detection}.
 However for flows dominated by turbulence rather than a large scale structure, the LCS manifolds form an ``intricate/complex tangle''  \citep{mathur2007uncovering} that is difficult to interpret.  
 The LCS is also limited in the time-scales it can reflect; the finite-times considered by the finite-time Lyapunov exponent calculation must be significantly shorter than the characteristic time of the flow (e.g. the eddy turn-over time, or the large-scale buoyancy time \citep{gibert_etal:T0, shearbursts} in a convective flow). This makes the LCS less useful for understanding an evolving turbulent flow where boundaries between regions can change rapidly.  Convex hull diagnostics provide a good alternative in this situation, since convex hulls evolve with the flow, and remain relevant as particles disperse over many characteristic time-scales.    Because LCS are closely tied to a short window of time, they roughly capture the major features of the initial velocity field.    This is demonstrated in the left panel of FIG.~\ref{lcscontourplot}, where maximal values of the finite-time Lyapunov exponent field are plotted as points over contours of the initial squared-velocity field.
  The right panel in FIG.~\ref{lcscontourplot} compares maximal values of the finite-time Lyapunov exponent field plotted with contours of convex hull volume.  
Results from the turbulent MHD convection simulation MC10 are shown in this figure because MC10 has the smallest Reynolds number of all simulations we produced.  The LCS results are therefore comparatively clear, and both LCS and convex hull results recognizably highlight the same complex structures.      Like Lagrangian coherent structures, convex hull diagnostics can provide information about regions with different dispersive behaviors; but because they employ more than two particles, convex hull results can be simpler to visually interpret for a turbulent flow.

\subsection{Intervals where convex hulls shrink}

\begin{figure}
\resizebox{6.5in}{!}{\includegraphics{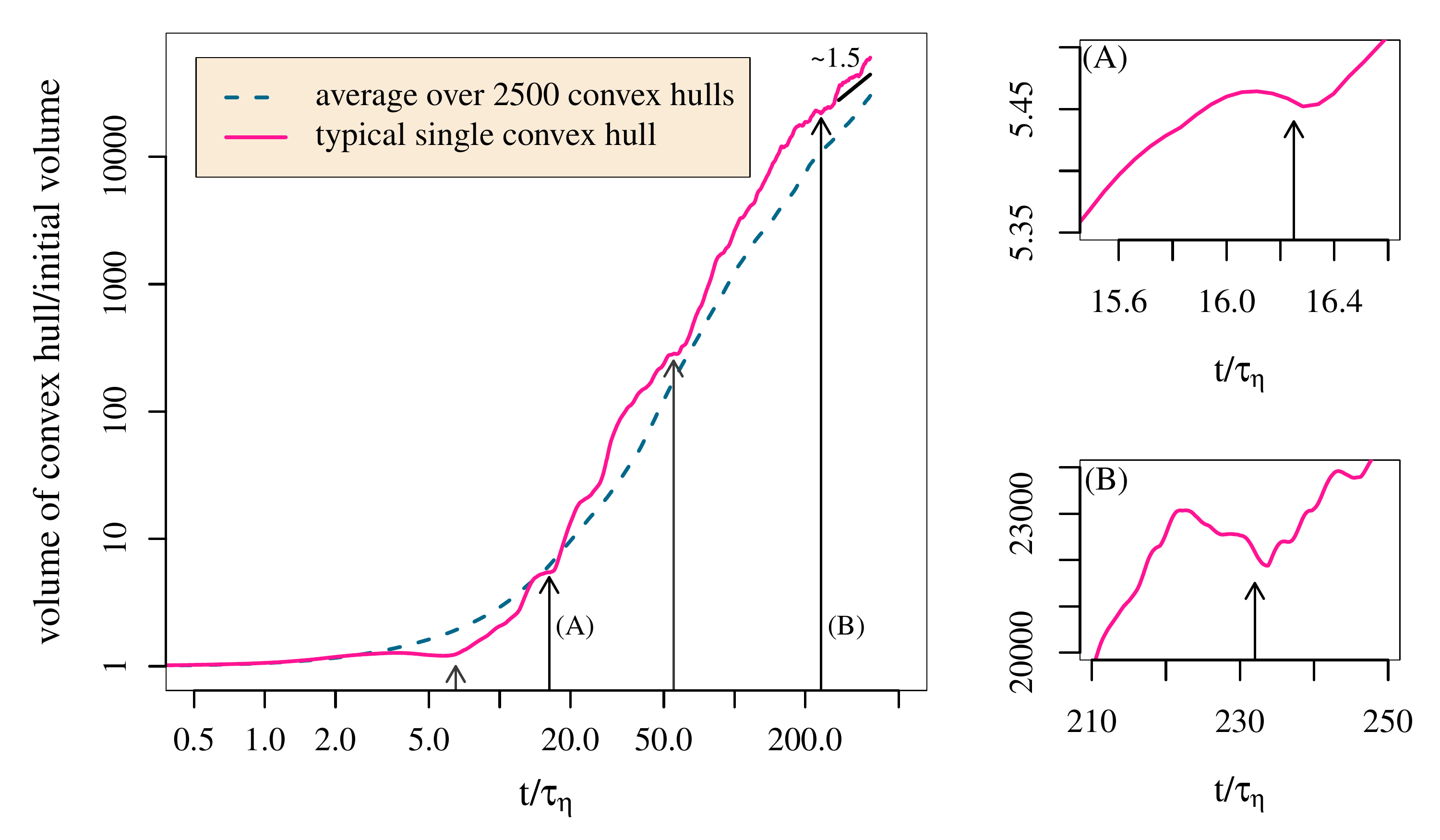}}
\caption{(Left) log-log plot of the evolution of the average volume of 2500 convex hulls compared with the volume of
a typical single convex hull for the simulation AMT.  The single convex hull volume selected contains 52 Lagrangian fluid particles.
Visible here are short
time-periods where the volume of the hull shrinks; four easily visible periods of volume shrinkage are marked by arrows.  
(Right) two short periods of volume shrinkage are shown in high detail in linear-scale plots. \label{wigxample} }
\end{figure}

Although convex hulls within any given simulation can grow at wildly different rates, they generally tend to expand as the particles that define them disperse.
Intriguingly, we find that all the simulations discussed here exhibit short episodes of time in which the surface area or
volume of a convex hull shrinks.  These periods
of shrinkage are not visible when an average over all convex hull volumes in a system is considered.  They are also absent from standard Lagrangian average particle-pair dispersion or average single-particle diffusion calculations.  The comparison between the average convex-hull volume and a single volume is shown in FIG.~\ref{wigxample}.  The dependence of shrinking episodes on the size of the initial hull and the number of particles was tested for simulation MC7 using large number of convex hulls of two different sizes, indicated in Table~\ref{simsuma}.  Periods of shrinkage occur smoothly, and independently of the number of Lagrangian fluid particles that are used to define the convex hulls -- they are present whether the groups of particles number in the 10s or 100s.  The shrinking episodes represent short periods where bunches of  tracer particles 
are dispersing in an anomalous way.

The relationship between convex-hull surface area and volume,
and how they are changing in time, is central to understanding how convex hull shrinkage can occur.  In many instances, the local flow is such that the volume decreases slightly while the surface area of the convex hull continues to grow.  This behavior reflects highly asymmetrical growth: the convex hull becomes stretched out in one direction, while becoming flatter in another.
However, we also observe other short periods of time when \emph{both} convex-hull surface area and volume \emph{decrease at the same time}.    This
can be explained by a scenario where some subset of particles that make up the convex hull are forced back toward each other, as would happen if a group of particles were confronted with a convergent flow.  We define these episodes by
\begin{eqnarray} \label{cond1}
\mathsf{volume}(t_i) &<& \mathsf{volume}(t_{i-1}) 
\\ \label{cond2}
\mathsf{surface~area}(t_i) &<& \mathsf{surface~area}(t_{i-1}) 
\end{eqnarray}
We identify intervals when a hull shrinks as intervals when conditions \ref{cond1} and  \ref{cond2}  are both satisfied.

Often when the surface area and volume simultaneously shrink, the volume shrinkage occurs over a slightly longer time-period.  This suggests that there can be a link between asymmetrical growth and a flow pattern that causes the convex hull to shrink.
If a convergent flow prevents particles from dispersing unhindered in one direction, the hull can continue to grow in other directions.  Thus the convex hull can grow in a highly-asymmetrical way for a short period of time, before the influence of the convergent flow is large enough to cause the surface area of the hull to shrink.  Indeed we find that shrinkage of convex hulls occurs most frequently when the physical system is subject to an asymmetry, i.e. under the influence of strong magnetic fields or vigorous convection.

\subsection{Time variation of convex hull shrinkage}

\begin{figure}
\resizebox{3.25in}{!}{\includegraphics{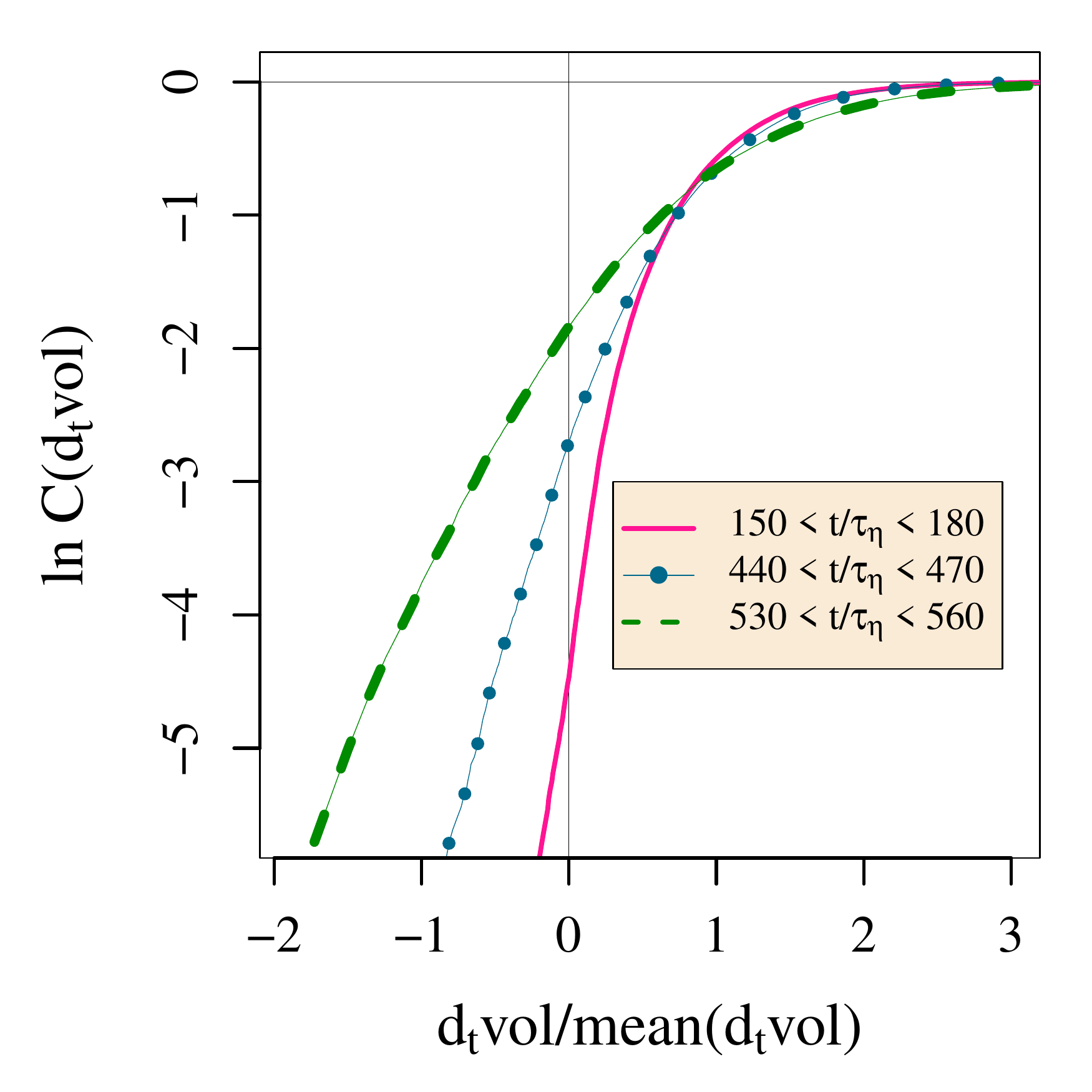}}\resizebox{3.25in}{!}{\includegraphics{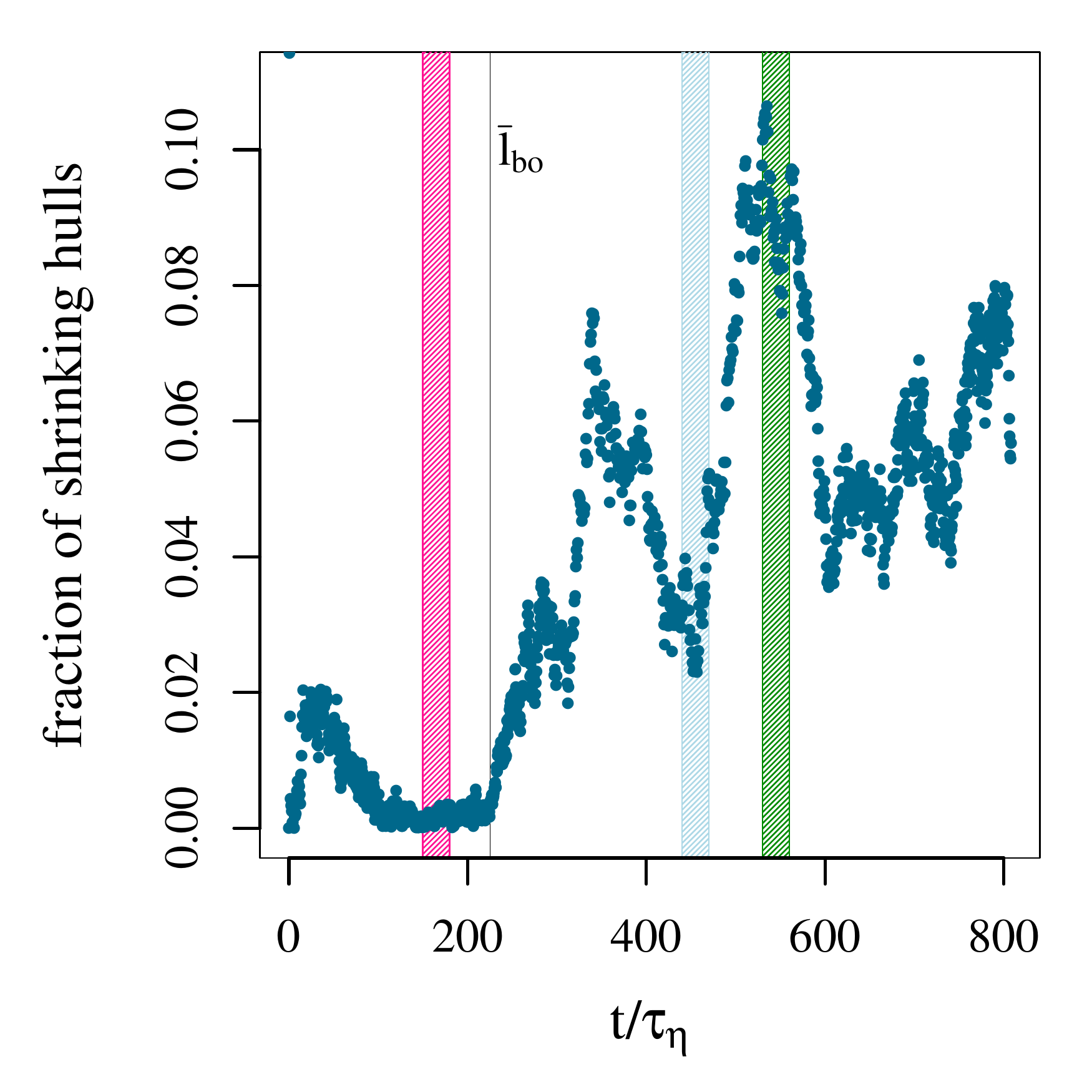}}
\caption{   (Left) cumulative distribution function of the time derivative of convex hull volume in simulation MC6 during 3 distinct equal-time periods highlighted in the right plot.  The value of the time derivative of volume is normalized to mean value of this derivative during each period.
 (Right)  time evolution of the fraction of hulls shrinking in simulation MC6 is shown.  A hull shrinks if surface area and volume at any instant in time are both less than their values at the previous time-interval.  Three equal time intervals are highlighted in order to closely examine the distribution of volume growth.  The time that the convex hull size matches the time-averaged Bolgiano-Obukhov length is indicated by a black line.
\label{convtimeconv} }
\end{figure}

We examine many convex hulls in each system, and the number of convex hulls that shrink at any point in time can vary greatly.
Because the particles contained in a convex hull tend to remain well-distributed and well-centered within the convex hull (as shown in FIG.~\ref{newclumpgraph}), this comparison between different points in time is reasonable. 
 The left panel of FIG.~\ref{convtimeconv} shows the cumulative distribution function of the instantaneous time derivative of convex hull volumes, examined during three equal-time periods in simulation MC6.   During extended periods in this convective flow, 10\% of the hulls in this simulation are shrinking on average.   This figure powerfully demonstrates that shrinkage plays a sufficiently large role to change the entire shape of the cumulative distribution function in a meaningful way.  Therefore shrinking is relevant to an over-all description of dispersion in this simulation.  At each point in time we examine a fraction of shrinking hulls, i.e. the number of hulls where both the surface area and volume are decreasing, divided by the total number of hulls in the system.   In the right panel of FIG.~\ref{convtimeconv}, the two largest peaks in the time-evolution of the fraction of shrinking hulls grow on a sufficiently short time-scale so that they are identifiable as a mild oscillation in the system-averaged convex hull volume.
 Comparing later times ($t> 600 ~\tau_\eta$) to earlier times
($t< 200 ~\tau_\eta$) in this plot, the average hull volume and surface area have increased by more than 5 times,
and the likelihood that a hull is shrinking has on average doubled.

To test the robustness of the results in the right panel of FIG.~\ref{convtimeconv}, we introduce a sliding threshold to filter out the smallest-scale changes in the convex hull volume and surface area.  In this test, we count a hull as shrinking if it satisfies both of the conditions:
\begin{eqnarray} \label{crit1}
\mathsf{volume}(t_i)  -  \mathsf{volume}(t_{i-1})   &<& -C ~\mathsf{volume}(t_0)
\\ \label{crit2}
\mathsf{surface~area}(t_i) - \mathsf{surface~area}(t_{i-1})   &<& -C~ \mathsf{surface~area}(t_0)
\end{eqnarray}
Eq. \ref{crit1} requires the decrease in convex hull volume in a time interval $t_i-t_{i-1}$ to be greater than a given fraction $C$ of the initial hull volume. 
Because a side of the initial hull volume spans more than 10 $\tau_\eta$, this is significantly larger than the smallest possible fluctuations.  
 Eq. \ref{crit2} expresses the same criterion as eq. \ref{crit1} for the surface area of the convex hull.  
For values of $C$ ranging from 0 -- 2 only minor changes are seen in the results of FIG.~\ref{convtimeconv}.  For $C=10$, the peaks in the right panel of FIG.~\ref{convtimeconv} drop by approximately half of their value.    The large fraction of convex hulls that we identify as shrinking are therefore not due to fluctuations of the convex hulls on the smallest spatial scales.

\begin{figure}
\resizebox{3.25in}{!}{\includegraphics{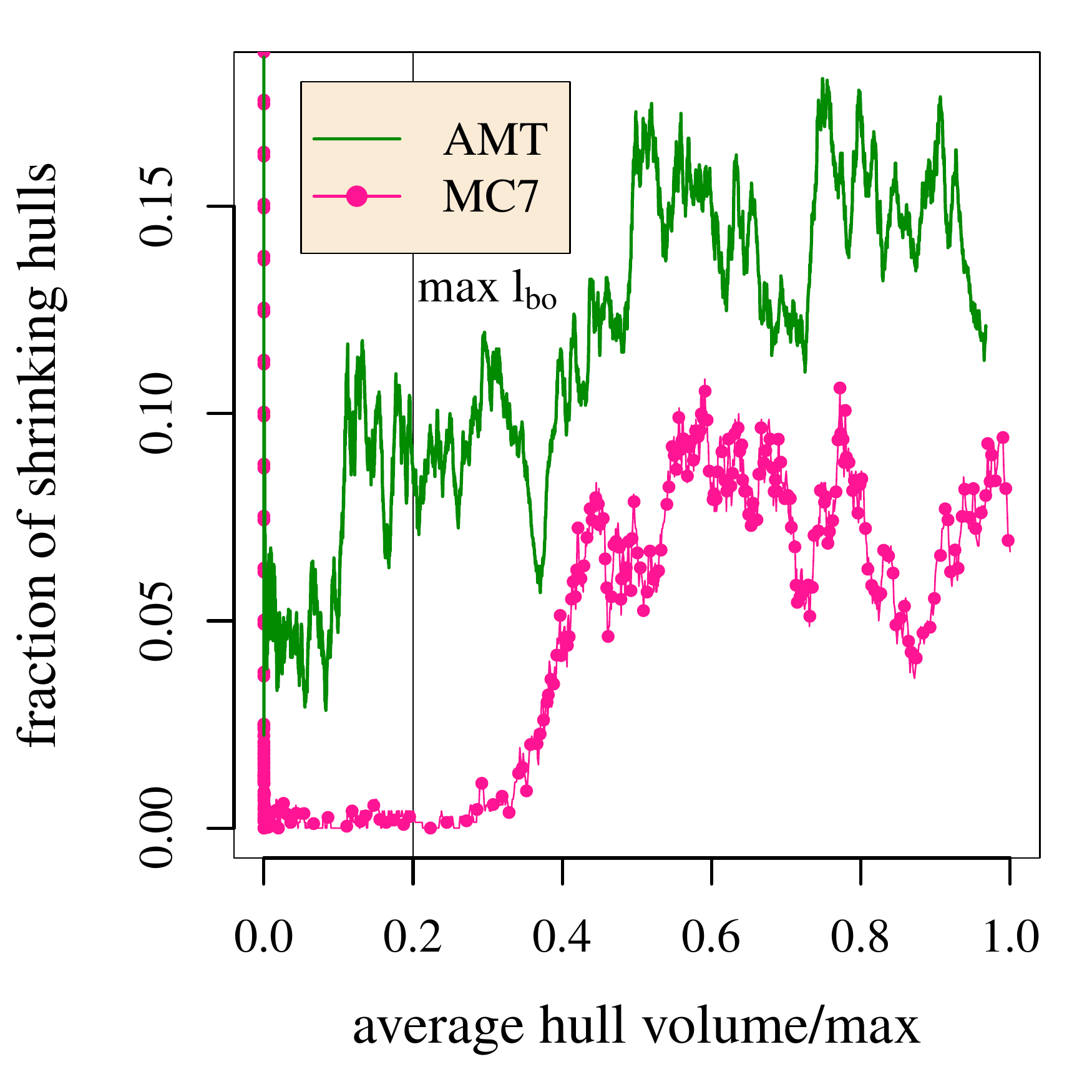}}\resizebox{3.25in}{!}{\includegraphics{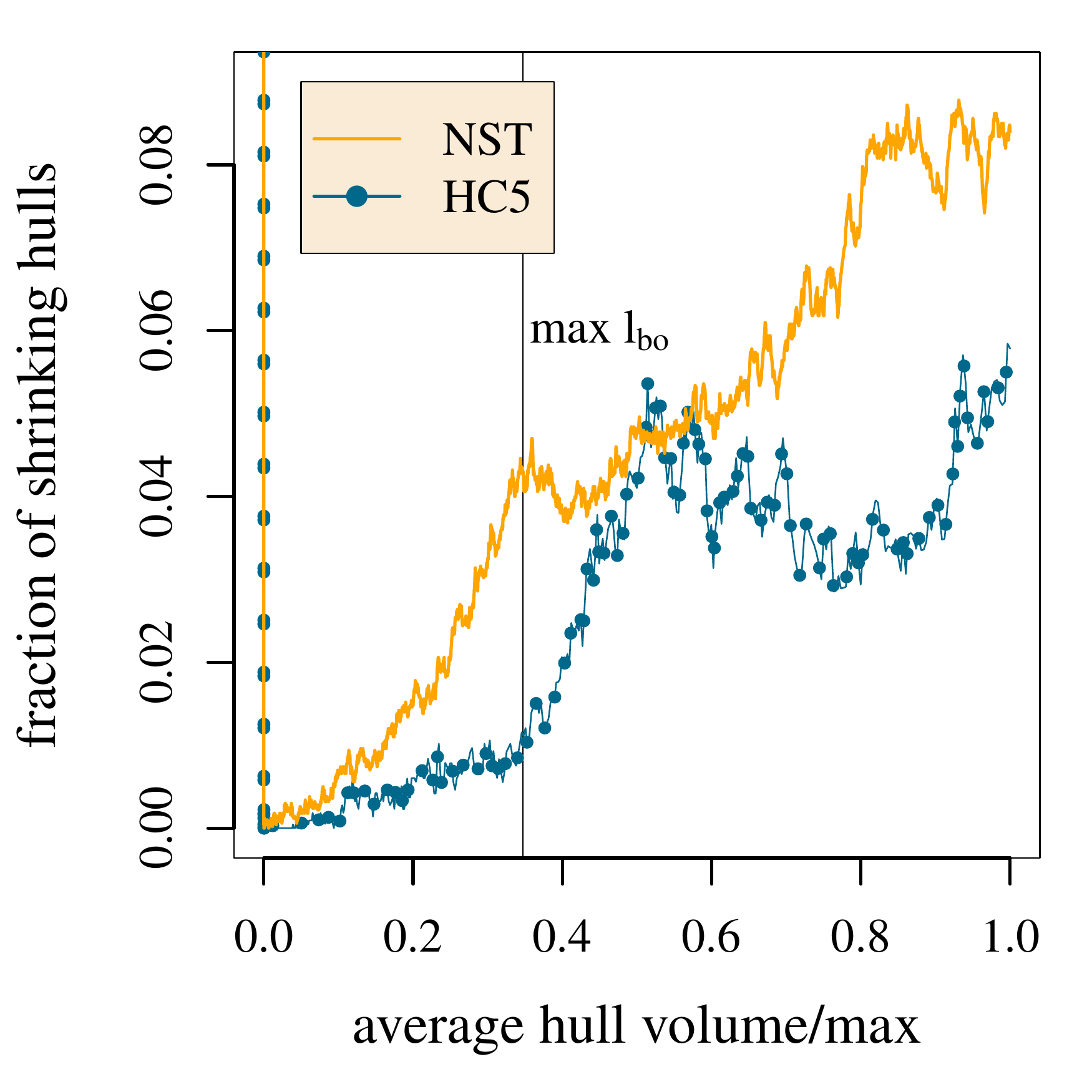}}
\caption{The growth of the fraction of hulls shrinking vs the average hull volume in the system, divided by its maximum,  is shown for (left) the anisotropic forced MHD simulation AMT and MHD convection simulation MC7, and (right) the forced turbulence simulation NST and hydrodynamic convection simulation HC5.
\label{convtimeturb} }
\end{figure}

In simulations where the turbulent flow is driven by random forcing, the fraction of shrinking hulls increases steadily with time, and thus also with the average size of the convex hulls, as is shown in FIG.~\ref{convtimeturb}.    For the forced turbulence simulation NST, the fraction of shrinking hulls grows smoothly without large oscillations.  For the forced anisotropic MHD simulation AMT,  the presence of a large mean magnetic field causes many more hulls to shrink simultaneously, and the time evolution of the fraction of shrinking hulls is noisier.  
Synchronicity of the shrinkage may be related to Alfv\'enic oscillations on scales comparable to the system size.

When the hulls are extremely small, during the first $\tau_{\eta}$ in any simulation, the convex hulls are likely to shrink small amounts at a rate above 5\%.  This is shown in the initial sharp drop in the curves in FIG.~\ref{convtimeturb}.  The simulations we examine span a range of Prandtl $\mathsf{Pr}$ and magnetic Prandtl $\mathsf{Pr_m}$ numbers.  Because different Prandtl numbers translate to different levels of small scale velocity fluctuations in turbulent convection, one might expect the Prandtl number to influence anomalous dispersion on small scales.  However we find no clear correlation between the size of the Prandtl and magnetic Prandtl numbers and the fraction of shrinking hulls during this initial period of dispersion.

In convection simulations a different pattern of time-evolution of the fraction of shrinking hulls emerges. The fraction remains uniformly small until the hulls in the system reach a critical size.  Typically this phase lasts more than a hundred Kolmogorov times.  This critical size is on the same scale as the Bolgiano-Obukhov length for each system.  An example of this is shown in  FIG.~\ref{convtimeconv} where the time when the time-averaged Bolgiano-Obukhov length  is reached is marked with a vertical black line.   Because scales above the Bolgiano-Obukhov length are convectively-driven, this implies that the periods of mass-shrinkage in convective simulations, i.e. where 5\% or more of the hulls are simultaneously shrinking, are influenced by large-scale convectively-driven flow structures.  

This provides us with a picture of the physics behind shrinkage in convecting simulations.   If a simple convective updraft encounters a much smaller group of particles, the particles will all be convected with the flow and no shrinkage will occur.  However, when a simple convective updraft encounters a group of particles that is roughly the same size, some of the Lagrangian particles that make up the hull can be temporarily pushed back toward the rest of the group.  Thus convection can result in fewer convex hulls shrinking when the hulls are small, and more shrinking when they are larger.

After hulls grow larger than the Bolgiano-Obukhov scale, the fraction of shrinking hulls grows to numbers comparable to the randomly forced simulations. The time evolution of fraction of hulls shrinking in the MHD convection simulations begins to oscillate as frequently as the forced MHD simulations.  We postulate that the frequent large-scale oscillations in this statistic signal Alfv\'en waves moving through the system.
Non-linear interactions in MHD simulations are more complex than in hydrodynamic simulations because of the presence of Alfv\'en waves on many time and length scales.
   In simulations of hydrodynamic convection, the fraction of shrinking hulls  grows comparatively more smoothly than in MHD simulations, although it still grows with more frequent oscillations than randomly forced turbulence.  This additional noisy growth in the fraction of  shrinking hulls is a consequence of the anisotropic flow and irregular convective drive.

\subsection{Variation of shrinkage across an ensemble of convex hulls}

\begin{figure}
\resizebox{3.25in}{!}{\includegraphics{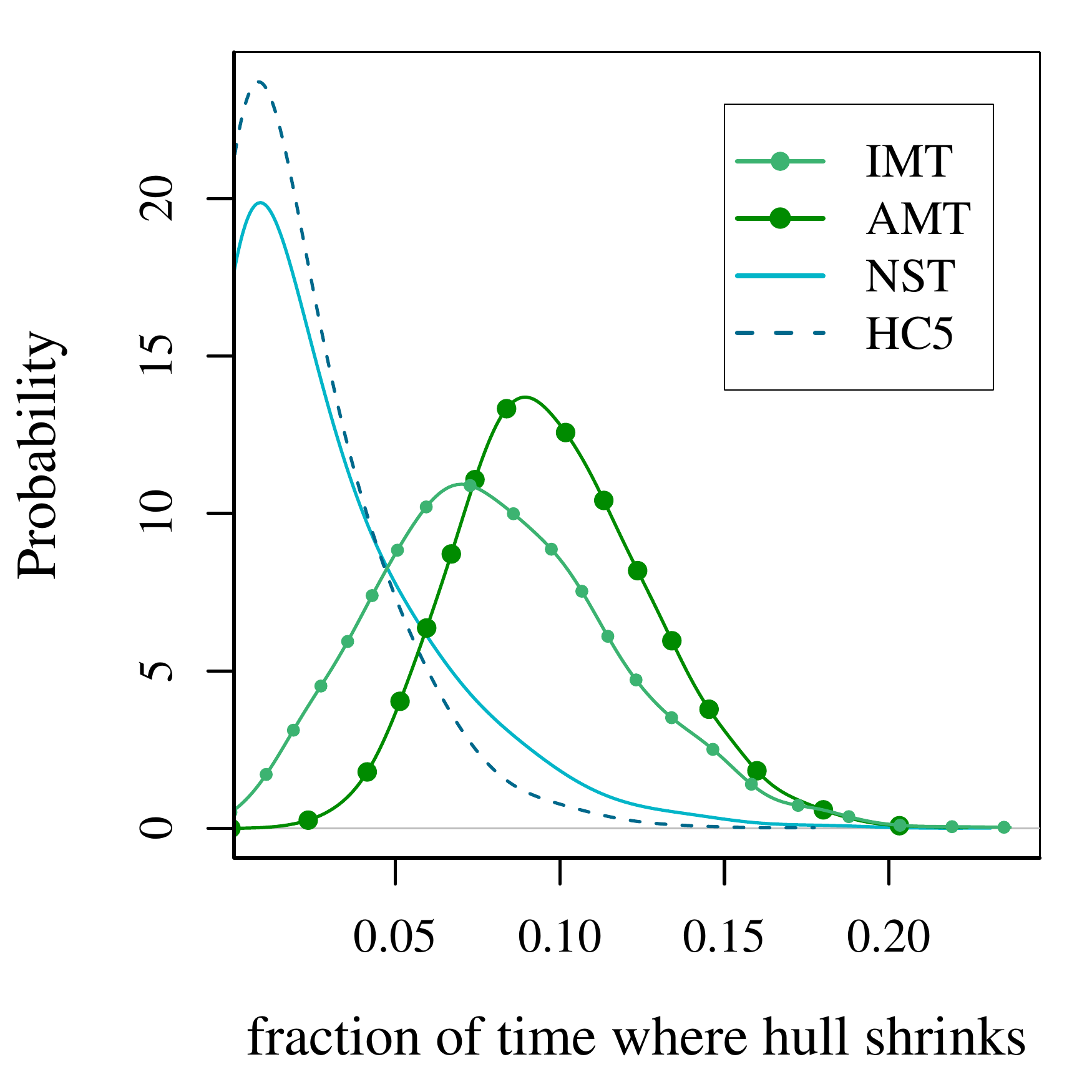}}\resizebox{3.25in}{!}{\includegraphics{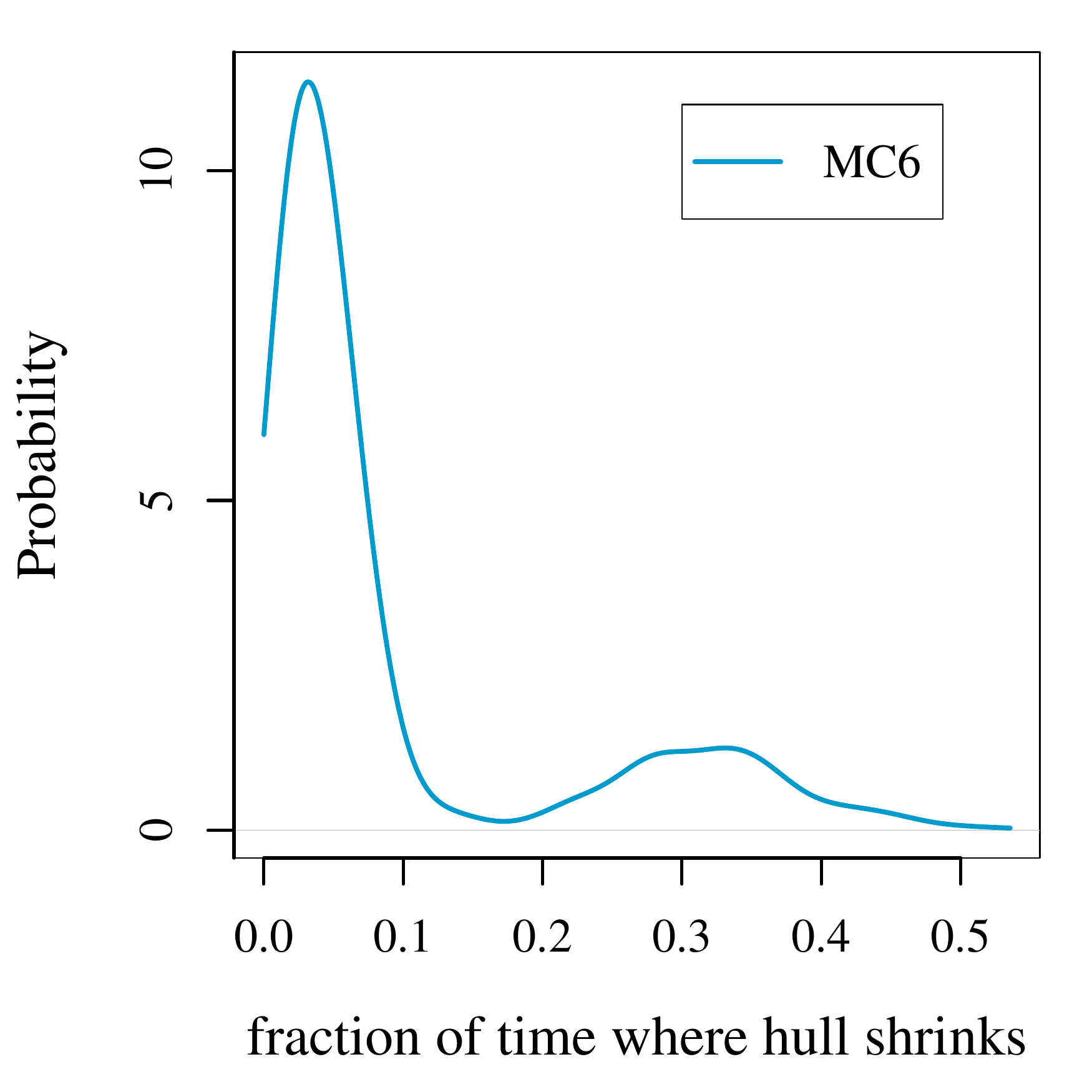}}
\caption{(Left) probability distribution of the fraction of Lagrangian crossing time that a convex hull is shrinking for the forced MHD simulations IMT and AMT, forced hydrodynamic turbulence simulation NST, and hydrodynamic convection simulation H5. The results from 5000 hulls are plotted for simulation NST, and 2500 hulls for AMT and IMT, 4420 hulls that span 9 horizontal slabs for HC5. (Right) probability distribution of the fraction of Lagrangian crossing time that each convex hull is shrinking for 3700 hulls that span 6 horizontal slabs in simulation MC6.
\label{shrinkhist} }
\end{figure}

To examine in detail the variation of convex hull shrinkage across an ensemble of convex hulls set in a regular spatial grid in the same simulation, we sum the amount of time
that each convex hull spends shrinking during the total time of the simulation.  The total time of each simulation is the Lagrangian crossing time.
The left panel in FIG.~\ref{shrinkhist} presents the probability density function of the fraction of time that a convex hull spends shrinking for simulations NST, IMT, AMT, and HC5. 
 For each
of our simulations, large numbers of hulls are examined so that the calculation of a probability density function is meaningful.
 In the randomly forced turbulence simulation NST, 16\% of convex hulls do not ever shrink.  Hulls that do shrink do so only occasionally; only 3.5\% of hulls shrink more than 10\% of the time.   The shapes of the curves in FIG.~\ref{shrinkhist} are impervious to our test for robustness expressed in eq. \ref{crit1} and \ref{crit2}.  With increasing values of $C$, slightly more hulls are counted as never shrinking, and slightly fewer hulls are counted as shrinking more than 10\% of the time.  Using a threshold of $C=0.1$ only 2.5\% of hulls shrink more than 10\% of the time in simulation NST.

    In simulation NST the probability density function peaks sharply for small values of the shrinking time and shows an approximately exponential
decay for higher shrinking times.
A similar shape is observed for simulations HC1, HC4, HC5, MC2, MC4, MC5, MC10, although the decay times vary between systems.
    In contrast, in the forced anisotropic MHD simulation AMT there are no hulls in the system that never shrink, and on average, hulls shrink about 10\% of the time.
The probability density function for this simulation has small values for small shrinking times, a broad peak at intermediate shrinking times, and decays for high shrinking times.
A similar profile shape is observed for simulations IMT, HC2, HC3, MC1, MC3, MC6, MC7, MC8, and MC9. However the two hydrodynamic
convection simulations HC2 and HC3 have comparatively low means, with convex hulls in these systems shrinking less than 5\% of the simulation time.
In FIG.~\ref{shrinkhist} the mean shrinking time in the anisotropic forced MHD simulation AMT is
significantly larger than in the isotropic forced MHD simulation IMT.  We conclude that the strength of the magnetic field has a strong influence on
anomalous diffusion and convex hull shrinkage.

Initial flow conditions are more important than physical parameters in determining the extent that shrinking will happen in a simulation; indeed simulations MC1 and MC5 have identical dissipative parameters, but the evolving flow in these simulations results in different characteristic probability density functions.   The probability density function for simulation MC6 is a particularly interesting case, with a heavy, possibly even bi-modal, right wing.  Some hulls in simulation MC6 are shrinking for more than half of the Lagrangian crossing time.  The extreme amount of shrinkage we see in MC6 correlates with the appearance of a relatively long-lived shear burst\citep{shearbursts}, a particular kind of localized coherent flow that involves high magnetic helicity, and growth of magnetic energy through stretching of magnetic field
lines.  The shear burst begins half way through simulation MC6, at about the time when the average hull size reaches the Bolgiano-Obukhov length.    The probability density function for MC6 is provided in the right panel of FIG.~\ref{shrinkhist}.  

The forced MHD simulations experience a higher number of shrinkage events than we observe during MHD convection.  These simulations have a 
significantly lower Alfv\'en ratio $r_{\mathrm{A}}$ than we were able to obtain from dynamo action during convection, and so experience higher magnetic activity.  
We postulate that with a lower Alfv\'en ratio, the frequency of shrinkage in MHD convection would be as high as that in forced MHD turbulence.

\section{Discussion}

We have developed the convex hull as a diagnostic for the intermittent nature of particle dispersion in turbulent flows and a general, versatile addition to traditional Lagrangian statistics.
Although the convex hull has been used to calculate volumes occupied by particles in some specialized contexts \citep{dietzel2013numerical, lakes}, this is the first time that the convex hull of the positions of Lagrangian tracer particles has been used as a fundamental diagnostic to obtain Lagrangian statistics of multi-particle dispersion in turbulent flows.  We examine large numbers of convex hulls of Lagrangian particles across a broad range of turbulence simulations that include magnetic fields and thermal convection.
The convex hull diagnostic may be useful in developing simple models \citep{pope2011simple} and a clearer over-all picture of particle dispersion in anisotropic and structured flows.
 The convex hull is designed to capture the extremes, rather than the means, of the excursions of a group of particles.  This type of diagnostic is of primary importance to studies of contaminants or energetic particles.  Convex hull
analysis may therefore be practical for analysis of Lagrangian tracer particles in oceanography and meteorology applications.   We use convex hull analysis to examine local regions of anomalous particle dispersion in turbulent flows structured by magnetic fields and convection in a way that is both intuitive and easy to visualize.
We find the convex hull to be a useful
alternative to Lagrangian coherent structure calculations when the Reynolds number of the simulation is high or long-term behavior is of interest.  

In analyzing patterns of dispersion, we find that convex hulls of groups of particles do not only grow in surface area and volume at different rates based on the conditions within a flow, but the hulls also sometimes shrink over short intervals.  
Because the particles that define the convex hull are expected to
separate on average, a short period where the convex hulls of large groups of particles do not grow, or indeed shrink, in both volume and surface area should be of great interest.  If a segment of the particles within a convex hull moves inward rather than outward, even if the overall change in volume and surface area is comparatively small, that is both unexpected and physically significant.
We focus in particular on these shrinking episodes, which we define by 
a simultaneous decrease in surface area and volume of a convex hull.  This situation of anomalous particle dispersion occurs sufficiently frequently to affect average statistical quantities.  Ten percent of the convex hulls shrink for ten percent of the simulation time on average in some MHD turbulence and in MHD convection simulations.
    
As convex hulls grow larger in size, they are more likely to experience periods of shrinking.  When convection drives the flow, the number of hulls that shrink increases rapidly once the size of the hulls reaches the scales that are convectively driven.  We conclude that convex hull shrinkage in convection simulations is often the result of interaction of a group of particles with a large-scale coherent flow, as when plumes are rising and sinking because of vigorous convection.

  Indeed convex hull shrinkage also escalates when large-scale Alfv\'en waves move with higher speed or higher amplitude through the system.   In simulation AMT a mean
magnetic field creates an anisotropy in the system, causing strong Alfv\'en waves at the lowest wave-number
modes to emerge.  Shrinkage happens significantly more often and more violently in this situation than in our isotropic MHD turbulence simulation.  Both of these randomly-forced MHD simulations exhibit more shrinkage than the homogeneous Navier-Stokes turbulence simulation, where no Alfv\'en waves are present.

\begin{acknowledgements}
{\small Dan Skandera collaborated to produce the data from simulation HC1.
This work has been supported by the Max-Planck
Society in the framework of the Inter-institutional
Research Initiative Turbulent transport and ion
heating, reconnection and electron acceleration in
solar and fusion plasmas of the MPI for Solar System
Research, Katlenburg-Lindau, and the Institute
for Plasma Physics, Garching (project MIFIF-A-AERO8047).  Simulations were performed on the VIP, VIZ, and HYDRA computer systems at the Rechenzentrum Garching of the Max Planck Society. }
\end{acknowledgements}

\bibliography{postdoc}

\end{document}